\documentclass[useAMS,usenatbib,a4paper]{mn2e}
\usepackage{amsmath}
\usepackage{amssymb}
\usepackage{amsfonts}
\usepackage {multirow}
\usepackage {graphicx}
\usepackage {natbib}
\usepackage{graphicx,amsmath}
\usepackage[hyperindex,breaklinks=true, colorlinks, citecolor=blue]{hyperref}  % for active links 
\usepackage{enumerate}
\usepackage{tablefootnote}
\usepackage{eurosym}
\usepackage[usenames,dvipsnames]{xcolor}

\def\reff@jnl#1{{\rm#1\/}}

\def\aj{\reff@jnl{AJ}}                  % Astronomical Journal
\def\araa{\reff@jnl{ARA\&A}}            % Annual Review of Astron and Astrophys
\def\apj{\reff@jnl{ApJ}}                % Astrophysical Journal
\def\apjl{\reff@jnl{ApJ}}               % Astrophysical Journal, Letters
\def\apjs{\reff@jnl{ApJS}}              % Astrophysical Journal, Supplement
\def\ao{\reff@jnl{Appl.Optics}}         % Applied Optics
\def\apss{\reff@jnl{Ap\&SS}}            % Astrophysics and Space Science
\def\aap{\reff@jnl{A\&A}}               % Astronomy and Astrophysics
\def\aapr{\reff@jnl{A\&A~Rev.}}         % Astronomy and Astrophysics Reviews
\def\aaps{\reff@jnl{A\&AS}}             % Astronomy and Astrophysics, Supplement
\def\azh{\reff@jnl{AZh}}                        % Astronomicheskii Zhurnal
\def\baas{\reff@jnl{BAAS}}              % Bulletin of the AAS
\def\jcap{\reff@jnl{JCAP}}              % Journal of Cosmology and Astroparticle Physics
\def\jrasc{\reff@jnl{JRASC}}            % Journal of the RAS of Canada
\def\memras{\reff@jnl{MmRAS}}           % Memoirs of the RAS
\def\mnras{\reff@jnl{MNRAS}}            % Monthly Notices of the RAS
\def\pra{\reff@jnl{Phys.Rev.A}}         % Physical Review A: General Physics
\def\prb{\reff@jnl{Phys.Rev.B}}         % Physical Review B: Solid State
\def\prc{\reff@jnl{Phys.Rev.C}}         % Physical Review C
\def\prd{\reff@jnl{Phys.Rev.D}}         % Physical Review D
\def\prl{\reff@jnl{Phys.Rev.Lett}}      % Physical Review Letters
\def\pasp{\reff@jnl{PASP}}              % Publications of the ASP
\def\pasj{\reff@jnl{PASJ}}              % Publications of the ASJ
\def\qjras{\reff@jnl{QJRAS}}            % Quarterly Journal of the RAS
\def\skytel{\reff@jnl{S\&T}}            % Sky and Telescope
\def\solphys{\reff@jnl{Solar~Phys.}}    % Solar Physics
\def\sovast{\reff@jnl{Soviet~Ast.}}     % Soviet Astronomy
 \def\ssr{\reff@jnl{Space~Sci.Rev.}}    % Space Science Reviews
\def\zap{\reff@jnl{ZAp}}                % Zeitschrift fuer Astrophysik
\def\nat{\reff@jnl{Nature}}             % Nature 

\def\ben{\begin{enumerate}}
\def\een{\end{enumerate}}
\def\bi{\begin{itemize}}
\def\ei{\end{itemize}}
\def\be{\begin{equation}}
\def\ee{\end{equation}}
\def\bea{\begin{eqnarray}}
\def\eea{\end{eqnarray}}
\def\ba{\begin{align}}
\def\ea{\end{align}}

\def\bdd{\boldsymbol{d }}

\def\bdm{\boldsymbol{m }}
\def\bdn{\boldsymbol{n }}
\def\bdN{\boldsymbol{N }}

\def\bdS{\boldsymbol{S }}
\def\bds{\boldsymbol{s }}

\makeatletter
\newcommand\footnoteref[1]{\protected@xdef\@thefnmark{\ref{#1}}\@footnotemark}
\makeatother

\title[Sensitivity/foreground modelling for CMB satellites]{Sensitivity and foreground modelling for large-scale CMB B-mode polarization satellite missions}
\author[M.\,Remazeilles et al.]{M.\,Remazeilles,$\!^{1}$\thanks{E-mail:~\url{mathieu.remazeilles@manchester.ac.uk}} C.\,Dickinson,$\!^{1}$\thanks{E-mail:~\url{clive.dickinson@manchester.ac.uk}} H.\,K.\,K.\,Eriksen,$\!^{2}$ I.\,K.\,Wehus,$\!^{3}$
\\
$^1$Jodrell Bank Centre for Astrophysics, Alan Turing Building, School of Physics and Astronomy, The University of Manchester, \\
Oxford Road, Manchester, M13 9PL, U.K.\\
$^2$Institute of Theoretical Astrophysics, University of Oslo, P.O. Box 1029 Blindern, NO-0315 Oslo, Norway\\
$^3$Jet Propulsion Laboratory, California Institute of Technology, Pasadena, CA 91109, USA\\
}

\begin{document}

\maketitle

%%%%%%%%%%%%%%%%%%%%%%%%%%%%%%%%%%%%%%%%%%%%%%%%%%%%
%%%
%%% Abstract
%%%
\begin{abstract}
The measurement of the large-scale B-mode polarization in the cosmic microwave background (CMB) is a fundamental goal of future CMB experiments. However, because of unprecedented sensitivity, future CMB experiments will be much more sensitive to any imperfect modelling of the Galactic foreground polarization in the reconstruction of the primordial B-mode signal. We compare the sensitivity to B-modes of different concepts of CMB satellite missions (LiteBIRD, COrE, COrE+, PRISM, EPIC, PIXIE) in the presence of Galactic foregrounds. In particular, we quantify the impact on the tensor-to-scalar parameter of incorrect foreground modelling in the component separation process. 
Using Bayesian fitting and Gibbs sampling, we perform the separation of the CMB and Galactic foreground B-modes. The recovered CMB B-mode power spectrum is used to compute the likelihood distribution of the tensor-to-scalar ratio. We focus the analysis to the very large angular scales that can be probed only by CMB space missions, i.e. the Reionization bump, where primordial B-modes dominate over spurious B-modes induced by gravitational lensing. 
We find that fitting a single modified blackbody component for thermal dust where the ``real'' sky consists of two dust components strongly bias the estimation of the tensor-to-scalar ratio by more than $5\sigma$ for the most sensitive experiments. Neglecting in the parametric model the curvature of the synchrotron spectral index may bias the estimated tensor-to-scalar ratio by more than $1\sigma$. For sensitive CMB experiments, omitting in the foreground modelling a $1\%$ polarized spinning dust component may induce a non-negligible bias in the estimated tensor-to-scalar ratio.
\end{abstract}


%%%%%%%%%%%%%%%%%%%%%%%%%%%%%%%%%%%%%%%%%%%%%%%%%%%%
%%%
%%% Intro
%%%
\section{Introduction}
\label{sec:introduction}

The detection of the CMB B-mode polarization, due to primordial gravitational waves, would provide a definitive proof that inflation occurred in the Early Universe. This is challenging because highly-polarized Galactic foregrounds dominate, by several orders of magnitude, the amplitude of the large-scale CMB B-modes. Therefore, the detection of the CMB B-modes is, by definition, a component separation problem \citep[][]{Betoule2009,Dunkley2009,Efstathiou2009,Bonaldi2011,Errard2011,Errard2012}.

\citet{Bicep2_2014} recently claimed the first detection of the primordial CMB B-modes at a tensor-to-scalar ratio $r=0.2$ in the high-latitude region of the sky observed by the ground-based CMB experiment {\it BICEP2}. The foreground subtraction performed in the {\it BICEP2} analysis was relying on poor assumptions on the polarized foreground model, which was constructed from the {\it Planck} intensity map of thermal dust \citep{Planck2013_XI}. More recently, the {\it Planck} Collaboration published their first polarized foreground data and showed that the {\it BICEP2} detection of B-modes was likely to be due to the polarized thermal dust emission \citep{Planck2015_Int_XXX}. In other words, the polarized dust signal in the {\it BICEP2} field was in first instance significantly under-estimated. The joint BICEP2/Keck and {\it Planck} analysis has now shown this to be the case with a combined upper limit of $r<0.11$ for a pivot scale of $k_0=0.002$\,Mpc$^{-1}$ \citep{BICEP_Planck2015}.

Due to the presence of the atmosphere, ground-based experiments, like {\it BICEP2}, are limited in frequency coverage. Furthermore, ground-based experiments are typically limited in terms of sky area coverage. This may generate larger uncertainties to the B-mode signal detected by ground-based experiments, limiting their ability to achieve robust foreground subtraction on large angular scales ($\gtrsim 1\degr$ or multipoles $\ell \lesssim 200$), where the cosmological signal from inflation is detectable. For example, the Reionization bump at $\ell \approx 10$, where the signal is due primarily to pure tensor perturbation modes. This is the reason why several fourth generation space-borne B-mode satellite concepts have emerged around the world (see Section~\ref{sec:experiments}). It is therefore highly likely that there will be at least one B-mode satellite in operation in the near future that will have the sensitivity to detect B-modes at the level of $r \sim 10^{-3}$ \citep[e.g. ][]{Matsumura2013}.

Component separation is clearly one of the biggest challenges for any B-mode satellite \citep[][]{Baccigalupi2004, Stivoli2006,Betoule2009,Dunkley2009,Bonaldi2011,Errard2012}. Most of the component separation methods rely on prior physical assumptions on foregrounds to clean the CMB (see \citet{Planck2013_XII} for a panel of methods used in the {\it Planck} temperature data analysis): {\tt Commander}  \citep{Eriksen2008} performs a Bayesian fitting based on a parametric model for foregrounds, {\tt SMICA} \citep[][]{Delabrouille2003, Cardoso2008} does not rely on a parametric model for foregrounds but still assumes a given number of independent foreground components to perform the spectral matching between the data and the model. However, the Galactic foreground B-modes, unlike the CMB, have an unknown spectral signature and, even worse, the exact number of independent polarized foregrounds is not known. More recent developments have been achieved in order to relax any prior assumption on the number of foregrounds, e.g. the {\tt GNILC} method \citep{Remazeilles2011b}.

Given the unprecedented sensitivity of future B-mode satellites, any slightly incorrect foreground modelling in the component separation process is expected to strongly bias the reconstruction of the CMB B-mode signal \citep{Armitage-Caplan2012}.

In this work we first compare the sensitivity of various CMB B-mode satellite concepts that have been put forward in recent years by evaluating the posterior probability distribution of the key cosmological parameter, the tensor-to-scalar ratio $r$.
We then go on to evaluate the impact of incorrect foreground modelling on the estimation of $r$. Incorrect modelling of foregrounds investigated in this work includes: i) fitting for a single spectral component of thermal dust where the simulated sky data contain a second spectral component of thermal dust not in the data model, ii) neglecting the curvature of the synchrotron spectral index in the data model, and iii) omitting the spinning dust polarization in the data model. Our aim is to achieve a comprehensive study by implementing a Bayesian fitting component separation method, {\tt Commander}, on a set of simulations with different foreground models and different instrument designs. Such a forecasting analysis is primordial for the optimisation of the concept of a future B-mode spatial mission. Note that in this comparative forecast analysis we do not consider the ability of each experiment to correct for lensing, limiting the analysis to low multipoles ($\ell < 12$) where lensing has negligible impact. In a realistic situation, space missions would also consider intermediate multipoles to guarantee a robust detection of the CMB B-modes, in which case lensing should be included for a fair comparison on larger range of multipoles, and the ability of high-resolution experiments to correct for lensing could make the difference. This should be investigated in a separate work.

In Section~\ref{sec:experiments} we summarise the main CMB B-mode satellites that have been proposed in recent times and tabulate their instruments in terms of frequency channels, beam sizes and sensitivities in each channel. Section~\ref{sec:models} discusses the sky models that are used in the simulations. Section~\ref{sec:method} summarises the method, based on the {\tt Commander} Bayesian parametric fitting code. Results are given in Section~\ref{sec:results} and conclusions summarised in Section~\ref{sec:conclusions}.

%%%%%%%%%%%%%%%%%%%%%%%%%%%%%%%%%%%%%%%%%%%%%%%%%%%%
%%%
%%% CMB satellite experiment concepts
%%%
\section{CMB B-mode satellite experiment concepts}
\label{sec:experiments}

\setlength{\tabcolsep}{0.5mm}

\begin{table*}
\scriptsize
  \centering
  \begin{tabular}{l|l|l|l|l|l}
\hline
Concept name   &Leading country/  &Frequencies       &Beam size        &Sensitivities           &Reference/notes \\
               &institution       &[GHz]             &FWHM [arcmin]    &[$\mu$K\,deg]                 &          \\
\hline \hline
EPIC-LC-TES    &U.S.A.            &30,40,60,90       &155,116,77,52,      &0.460,0.156,0.085,0.037,        &EPIC Low-Cost option with\\
               &(NASA)            &135,200,300       &34,23,16      &0.035,0.037,0.062  &TES detectors \citep{Bock2008}   \\
EPIC-CS        &U.S.A.            &30,45,70,100,       &15.5,10.3,6.6,4.6,      &    0.683,0.367,0.150,0.117    &EPIC Comprehensive-Science option \\
               &(NASA)            & 150,220,340,500       &3.1,2.1,1.4,0.9      &0.117,0.183,0.883,7.50  & \citep{Bock2008}   \\
EPIC-IM-4K     &U.S.A.            &30,45,70,100,150  &28,19,12,8.4,5.6 &0.147,0.061,0.027,0.018,0.014, &EPIC Intermediate with \\
               &(NASA)            &220,340,500,850   &3.8,2.5,1.7,1.0  &0.027,0.058,0.014,0.012  & 4\,K mirror \citep{Bock2009} \\
\hline
LiteBIRD       &Japan             &60,78,100,        &75,58,45,        &0.172,0.108,0.078,     & \citep{Matsumura2013}\\
               &(JAXA)            &140,195,280       &32,24,16         &0.062,0.0517,0.063      &                      \\
\hline
COrE          &Europe            &45,75,105,135,165, &23.3,14,10,7.8,6.4,  &0.150,0.078,0.077,0.075,0.077 &ESA M mission concept \\  
               &(ESA)             &195,225,255,285,315,&5.4,4.7,4.1,3.7,3.3,&0.075,0.075,0.173,0.283,0.767, &\citep{Core2011} \\
               &                  &375,435,555,675,795 &2.8,2.4,1.9,1.6,1.3 &1.95,4.25,9.82,57.0,348.0  &            \\
\hline
COrE+          &Europe            &60,70,80,90,100, &21.0,18.0,15.8,14.0,12.6,  &0.485,0.467,0.320,0.257,0.197,  & ESA M mission concept 3\tablefootnote{\label{foot1}\url{http://conservancy.umn.edu/handle/11299/169642}
}\\    
  Light                   &(ESA)             &115,130,145,160,175,&11.0,9.7,8.7,7.9,7.2,&0.138,0.110,0.092,0.092,0.090,  & \\     
               &                  &195,220,255,295,340, &6.5,5.7,5.0,4.3,3.7, & 0.090,0.135,0.218,0.430,0.817, &            \\
          &                  &390,450,520,600 & 3.2,2.8,2.4,2.1 &1.645,4.205,10.535,15.848  &            \\

COrE+          &Europe            &60,70,80,90,100, &14.0,12.0,10.5,9.3,8.4,  &0.342,0.233,0.160,0.123,0.098, & ESA M mission concept 4$^{\ref{foot1}}$\\    
Extended       &(ESA)             &115,130,145,160,175,&7.3,6.5,5.8,5.3,4.8, &0.073,0.057,0.057,0.057,0.058,  & \\     
               &                  &195,220,255,295,340, &4.3,3.8,3.3,2.9,2.5, &0.063,0.090,0.152,0.220,0.422,  &            \\
              &                   & 390,450,520,600,700,800 &2.2,1.9,1.6,1.4,1.2, 1.1&   0.790,1.982,5.632,20.05,93.5,203   & \\   
\hline
PRISM          &Europe            &30,36,43,51,62,    &17,14,12,10,8.2, &0.211,0.141,0.133,0.103,0.098,  &ESA L mission concept   \\              
               &(ESA)             &75,90,105,135,160, &6.8,5.7,4.8,3.8,3.2 &0.093,0.078,0.068,0.061,0.0572 &\citep{Prism2014} \\
               &        &185,200,220,265,300, &2.8,2.5,2.3,1.9,1.7, &0.059,0.061,0.064,0.073,0.085,   &                  \\
               &                  &320,395,460,555,660  &1.6,1.3,1.1,0.92,0.77 &0.092,0.135,0.197,0.404,0.953   &                 \\
\hline
PIXIE         &U.S.A.            &   30,60,90,120,150,               & 96.0 (constant)           &       5.180,1.390,0.691,0.454,0.352,                    &      \citep{Pixie2011}           \\
               &(NASA)           &   180,210,240,270,300,        &                &         0.307,0.292,0.297,0.319,0.358              &                 \\
              &           &   330,360,390,420,450,               &                &          0.418,0.503,0.623,0.790,1.020,        &                 \\
             &           &   480,510,540,570,600,               &                 &          1.350,1.800,2.440,3.350,4.660,        &                 \\
            &           &   630,660,690,720,750,               &                  &    6.550,9.280,13.30,19.10,27.70,      &                 \\
           &           &    780,810,840,870,900,              &                   &        40.50,59.60,88.20,131.00,196.00,      &                 \\
          &           &      930,960,990,1020,1050,            &                  &        294.00,444.00,672.00,1020,1560,    &                 \\
          &           &      1080,1110,1140,1170,1200            &                &         2390,3670,5660,8750,13600   &                 \\
\hline
  \end{tabular}
  \caption{Summary of proposed CMB B-mode satellite missions. The frequencies, beam sizes, and sensitivities are the nominal values for each of the missions (see text). The sensitivities are to Stokes $Q$ (or $U$) and are given in CMB thermodynamic units.}
  \label{tab:experiments}
\end{table*}

\setlength{\tabcolsep}{2mm}

A number of concepts for CMB B-mode satellite missions have emerged over recent years, particularly after the success of {\it COBE} and {\it WMAP}, and more recently, {\it Planck} \citep{Planck2011_I}. These proposed missions are focussed on measuring the CMB polarization across a wide range of angular scales to an accuracy that is limited by astrophysical foregrounds. The sensitivities to the CMB B-modes are at least 2 orders of magnitude better than the current general of ground- and balloon-based experiments, which would allow a $r\sim0.001$ to be detected.  Each concept is briefly described below. Table~\ref{tab:experiments} gives the basic experimental set-up for each concept, based on the designs described in the literature.

\subsection{EPIC}

 EPIC, Experimental Probe of Inflationary Cosmology \citep{Bock2008,Bock2009}, is a U.S. B-mode satellite mission concept proposed to the National Aeronautics and Space Administration (NASA). The EPIC study team has investigated three possible designs with different instrumental characteristics.

\subsubsection{EPIC Low-Cost option with TES detectors}

The Low-Cost option \citep{Bock2008}, termed EPIC-LC, consists of six $30$\,cm refracting telescopes for an optimisation of the cost of the mission, large arrays of Transition-Edge Superconducting (TES) bolometers, and a rotating half-wave plate in front of the optics to modulate the polarization signal without modulating any polarization systematics generated from the optics. It has a limited resolution of between $16$\,arcmin and $155$\,arcmin and $7$ frequency channels distributed between $30$\,GHz and $300$\,GHz. EPIC-LC is therefore focussed on the large-scale primordial CMB B-mode science. Two types of detectors were considered: i) Neutron-Transmutation-Doped (NTD) Ge Transition Edge Sensor (TES) bolometers with unmultiplexed JFET amplifiers (the same technology developed for the {\it Planck} HFI), or ii) improved SQUID-multiplexed TES bolometers, which provide higher sensitivity. Here, we only provide results for the higher sensitivity TES option.

\subsubsection{EPIC Comprehensive-Science option}

Comprehensive-Science option \citep{Bock2008}, termed EPIC-CS, uses a 3\,m diameter dish, which guarantees access to a broader CMB science case compared to EPIC-LC. EPIC-CS has a higher resolution ($\approx 5$\,arcmin in the CMB channel and up to $1$\,arcmin at high frequency) and a broader frequency range of observation ($8$ channels from $30$\,GHz to $500$\,GHz) than the low-cost option. However, EPIC-CS has a lower sensitivity ($0.12\,\mu$K\,deg at $150$\,GHz) than EPIC-LC ($0.03\,\mu$K\,deg at $135$\,GHz).

\subsubsection{EPIC Intermediate option with 4\,K mirror}

The Intermediate option \citep{Bock2009}, termed EPIC-IM, consists of an intermediate size telescope ($1.4$\,m diameter) cooled to $4$\,K to reduce thermal noise. The angular resolution of EPIC-IM is similar to that of EPIC-CS, but the sensitivity of EPIC-IM is higher due to a mechanical $4$\,K cooling system and a focal plane of detectors designed in hexagonal cells. Moreover, EPIC-IM can observe the sky through a larger frequency range than the two other options with $9$ channels distributed between $30$\,GHz and $850$\,GHz. There was also a less expensive option with a 30\,K mirror, which we do not consider here.

\subsection{LiteBIRD}

LiteBIRD,  Lite (Light) satellite for the studies of B-mode polarization and Inflation from cosmic background Radiation Detection \citep{Matsumura2013}, is a Japanese satellite mission concept entirely dedicated to the detection of the large-scale primordial CMB B-mode polarization. This space mission is led by the Japan Aerospace Exploration Agency (JAXA). It was selected in March 2014 as one of $27$ highest-priority large projects by the Science Council of Japan. 

LiteBIRD is a light satellite with two small mirrors of $\approx 60$\,cm diameter cooled to $4$\,K, including 2000 TES (or MKIDs) bolometric detectors, and a half-wave plate in front of the instrument in order to modulate the polarized signal and to reduce the systematics. LiteBIRD should observe the sky within six frequency bands between $60$\,GHz and $280$\,GHz with a typical sensitivity of $0.06\,\mu$K\,deg at the CMB frequency channel ($140$\,GHz). 

Being focused on the detection of large-scale CMB B-modes, LiteBIRD does not rely on high-resolution channel observations. The highest resolution is reached at the $280$\,GHz channel with a $16$\,arcmin beam full width at half maximum (FWHM).

\subsection{COrE}

COrE, Cosmic Origin Explorer \citep{Core2011} is a European B-mode satellite concept proposed to the European Space Agency (ESA) in 2011 as a medium-class (M) mission. 

The design of COrE allows for observing the microwave sky in polarization over $15$ frequency bands between $45$\,GHz and $795$\,GHz. While the COrE experiment provides an angular resolution comparable to {\it Planck}, the sensitivity of COrE is $10$ to $30$ times better than {\it Planck} because of a hundred times more detectors. COrE includes a rotating half-wave plate in order to better control the systematics (in particular temperature leakage) when measuring the polarization modes. Due to the presence of the half-wave plate, the size of the antenna can not be too large ($\approx 1.5$\,m) in order to meet the size requirements of a M-class spatial mission.

\subsection{COrE+}

COrE+, Cosmic Origin Explorer Plus (see Table~\ref{tab:experiments} for reference) is a European B-mode satellite concept proposed to ESA in 2015 as an M-class mission. It was a follow-on from the original COrE concept. Two different designs have been investigated: a low-cost light satellite and an extended version with a larger telescope mirror.

\subsubsection{COrE+ Light concept}

In the COrE+ Light concept, the telescope has a $1$\,m aperture allowing for $6$\,arcmin resolution at $200$\,GHz. COrE+ Light can observe the sky through $19$ frequency channels between $30$\,GHz and $600$\,GHz with a claimed sensitivity of $2.5\,\mu$K\,arcmin in CMB polarization after foreground cleaning. COrE+ Light is the European low-cost ($\approx$ \euro550M) and low-risk option. 

\subsubsection{COrE+ Extended concept}

The COrE+ Extended version has a $2$\,m aperture telescope, guaranteeing $\approx 4$\,arcmin resolution at $200$\,GHz, and an extended frequency range of observations with $21$ channels distributed between $30$\,GHz and $800$\,GHz. The Extended version is expected to provide $1.5\,\mu$K\,arcmin sensitivity in CMB polarization after foreground cleaning. The Extended mission was proposed to be funded with ESA and international partners, at a level of $\approx$ \euro700M. 

\subsection{PRISM}

PRISM, Polarized Radiation Imaging and Spectroscopy Mission \citep{Prism2014}, is a European space mission concept proposed to the European Space Agency (ESA) in 2013 as a large-class mission for characterizing the microwave sky in intensity and polarization. 

 With a relatively high resolution (few arcmin to $1$ arcmin), a high sensitivity (between $0.05\,\mu$K\,deg and $0.9\,\mu$K\,deg), and a large frequency range through $20$ observation channels distributed from $30$\,GHz to $660$\,GHz, PRISM is designed to cover a broad range of CMB science. This includes CMB B-mode search, Sunyaev-Zeldovich cluster science, gravitational lensing, Cosmic Infrared Background, Galactic Science, and spectral distortions of the CMB absolute spectrum. PRISM is considered as the ultimate CMB mission in terms of sensitivity and resolution. 

PRISM consists in two instruments: a high angular resolution polarimetric imager with a $3.5$\,m diameter telescope cooled to $\approx$\, 4 K, and a low angular resolution spectrometer to measure the spectral distortions to the absolute black-body spectrum of the CMB.  For this study, we only simulate data from the imaging instrument.

\subsection{PIXIE}

The Primordial Inflation Explorer (PIXIE) is a U.S. space mission concept proposed to NASA, consisting of a polarizing Fourier transform spectrometer (FTS) with 400 spectral bands from $30$\,GHz to $6000$\,GHz \citep{Pixie2011}. Despite the fact that the instrument is designed to perform absolute spectroscopy, it can also be exploited to measure large-scale CMB B-mode polarization. The FTS operates as a nulling polarimeter by cancelling any unpolarized emission. PIXIE channels have approximately the same, relatively low, angular resolution of $\approx 96$\,arcmin beam FWHM.  The map sensitivity in each channel can be varied depending on the operation of the FTS, resulting in variable bandwidths. For this study, we use sensitivities appropriate for 40 channels between 30 and 1200\,GHz (A. Kogut, priv. comm.).

%%%%%%%%%%%%%%%%%%%%%%%%%%%%%%%%%%%%%%%%%%%%%%%%%%%%
%%%
%%% Sky model simulation set
%%%
\section{Sky models and simulations}
\label{sec:models}

We consider a set of polarized sky simulations generated using version 1.7.8 of the Planck Sky Model ({\tt PSM}) software package\tablefootnote{\url{http://www.apc.univ-paris7.fr/~delabrou/PSM/psm.html}} \citep{Delabrouille2013}. The software is also used to simulate Gaussian white noise at each frequency for each experiment, using the noise levels and angular resolutions given in Table~\ref{tab:experiments}. No bandpass effects are included, such that each frequency is assumed to be monochromatic. 

\subsection{CMB}

The CMB is chosen to be a Gaussian field simulated from an input power spectrum defined by a set of standard cosmological parameters based ($\Omega_b=0.0456$, $\Omega_m=0.272$, $\Omega_k=0.0$, $H_0=70.4$\,km\,s$^{-1}$\,Mpc$^{-1}$, $\sigma_8=0.809$, $A_s=2.441 \times 10^{-9}$, $n_s=0.963$, $\tau=0.087$) based on recent data\footnote{Note that during the last stages of the completion of this work, new results from the  observations of the Planck satellite have been posted, pointing to a lower value of the optical depth to Reionization with $\tau=0.066$ \citep{2015arXiv150201589P}. This means that for a given tensor-to-scalar ratio, the amplitude of CMB B-modes with respect to foreground B-modes should be even lower.}. We use three values of the tensor-to-scalar ratio, $r$, to test the effects of sample variance: 0.0 (no B-modes therefore no sample variance), 0.001 (minimal sample variance, close to the sensitivity level of the experiments we are testing) and 0.05 (just below the current upper limits of $r<0.1$; \citealt{BICEP_Planck2015}). Note that we use a pivot scale of $k_0=0.05$\,Mpc$^{-1}$ but the results are very weakly dependent on this because we have no tilt in the tensor spectrum ($n_T=0.0$). The input theoretical power spectrum is then calculated using the {\tt CAMB} software \citep{Lewis2000}. We do not include the effects of gravitational lensing, which would contribute significant power at scales $\ell \gtrsim 100$. For a fair comparison, we use the same realization of the CMB sky for each experiment and model. The main parameter that we vary is the tensor-to-scalar ratio, $r$. The CMB fluctuations in both intensity and polarization are scaled with frequency according to the same spectral emission law, $a(\nu)$, given by the derivative with respect to temperature of a blackbody spectrum at $T_{\rm CMB} = 2.725$\,K:   
\bea
a(\nu) =  {d\,B_{\nu}(T)\over d\,T}\bigg\vert_{T=T_{\rm CMB}}.
\eea

\subsection{Foregrounds}

The simulation of the polarized sky includes the emission from Galactic foregrounds: synchrotron radiation and thermal dust radiation. For specific purposes in the present work, we also include in some simulations of the set the polarization radiation from a spinning dust component. We do not include the effects of polarized sources, since they are only dominant on smaller angular scales ($\ell \gtrsim 100$). 

\subsubsection{Synchrotron}

For the synchrotron component, the $Q^{sync}_\nu$ and $U^{sync}_\nu$ polarization maps are simulated at different frequencies by extrapolating the {\it WMAP} polarization observations at $23$\,GHz \citep{Kogut2007} through a simple power-law 
\bea
Q^{sync}_\nu &=& Q_{23\, GHz}\left({\nu\over 23}\right)^{\beta_s}, \cr
U^{sync}_\nu &=& U_{23\, GHz}\left({\nu\over 23}\right)^{\beta_s},
\eea
considering either a uniform spectral index, ${\beta_s = -3}$, or a non-uniform synchrotron spectral index \citep{Miville-Deschenes2008} over the sky. We also consider for some of the simulations the possibility that the synchrotron spectral index includes a frequency dependence, i.e.
\bea
\beta_s (\nu) = -3 + C\log\left(\nu\over 23\right),
\label{eq:curvature}
\eea
through the addition of a curvature term, $C$. We adopt values of $C=0.3$ and $C=-0.3$, resulting in a flattening or steepening synchrotron spectrum at frequencies above 23\,GHz \citep{Kogut2007}. 

\subsubsection{Thermal dust}

For the thermal dust, the $Q^{dust}_\nu$ and $U^{dust}_\nu$ polarization maps are derived from the intensity map $I_{\nu}$ as
\bea
Q^{dust}_\nu &=& f_d\,g_d I_{\nu} \cos\left(2\gamma_d\right), \cr
U^{dust}_\nu &=& f_d\,g_d I_{\nu} \sin\left(2\gamma_d\right).
\eea
The dust polarization angle map, $\gamma_d$, and the geometric depolarization map, $g_d$, due to magnetic field configuration \citep{Miville-Deschenes2008}, are coherent with those of the polarized synchrotron model.
The dust polarization factor is set to $f_d=0.15$, which, after correction by the geometric depolarization factor $g_d$, results in an observed polarization fraction, $f_d\,g_d$, of about $5$\,\% on average over the sky. This is slightly lower than the typical value at high latitudes ($\approx 8$\,\%) that has recently been observed by the {\it Planck} satellite at 353\,GHz \citep{Planck2015_Int_XIX,Planck2015_IX}; the precise values depend on the zero levels in the {\it Planck} data and the polarization fraction varies significantly over the sky up to a maximum of $\approx 20\,\%$ along some sight-lines.

For the dust intensity map, $I_\nu$, we choose to extrapolate at different frequencies either the {\it Planck} thermal dust opacity map at $353$\,GHz \citep{Planck2013_XI} or the Schlegel-Finkbeiner-Davis (SFD98) dust intensity map at $100$ micron \citep{SFD1998}. We consider two alternative models of emission law for extrapolating the dust template over frequencies: either a single modified blackbody (MBB) spectrum (Model 1a and Model 2a in Table~\ref{tab:models})
\bea
I_{\nu} = I_{353} \left({\nu\over 353}\right)^{\beta_d} B_{\nu}(T_d), 
\eea
with uniform dust spectral index $\beta_d=1.6$ and uniform dust temperature $T_d=18$\,K, in which case we extrapolate the {\it Planck} dust model at $353$\,GHz, $I_{353}$, or a more complex model of emission (Model 1b and Model 2b in Table~\ref{tab:models}) parametrized by two modified blackbodies based on model 7 of \cite{Finkbeiner1999}: 
\bea
I_{\nu} = I_{100\mu m} \left[f_1\left({\nu\over 3000}\right)^{\beta_1} B_{\nu}(T_1)+f_2\left({\nu\over 3000}\right)^{\beta_2} B_{\nu}(T_2)\right],
\eea
accounting for different populations of dust grains: a fraction $f_1$ of cold dust component with temperature $T_1 = 9.6$\,K and spectral index $\beta_1 = 1.5$, and a dominant fraction $f_2$ of hot dust component with temperature $T_2 = 16.4$\,K and spectral index $\beta_2 = 2.6$. In this case, the SFD98 dust intensity map at $100$ micron, $I_{100\mu m}$, is extrapolated over frequencies. We note that recent results from the {\it Planck} satellite suggest that the Rayleigh-Jeans thermal dust spectral index varies relatively little over the sky, with an average value of $\beta_d=1.59 \pm 0.02$ and a $1\sigma$ dispersion of 0.16 \citep{Planck2015_Int_XXII}. Therefore our model is arguably more complex than the real sky. However, this is partly compensated by the fact that we are using a smaller average polarization fraction than observed. These details do not change the major conclusions of this paper.

\subsubsection{Spinning dust}

For simulations Model 1f and Model 2f (Table~\ref{tab:models}), we also include a spinning dust component with $1\%$ fraction of polarization \citep{Dickinson2011,Rubino-Martin2012} and same polarization angles than thermal dust. The intensity map for the spinning dust component is taken from the {\it Planck} thermal dust map at $353$\,GHz \citep{Planck2013_XI}, rescaled by $0.91$\,K/K using the correlation coefficient from \cite{Planck2015_XXV}, and extrapolated at each frequency from the $23$\,GHz value assuming a cold neutral medium (CNM) model of emission law \citep{Ali-Haimoud2009}.

\setlength{\tabcolsep}{1mm}  % reduce table column separation by half

\begin{table}
\scriptsize
  \centering
  \begin{tabular}{l|l|l|l|l}
\hline
Model ID   & CMB  & Synchrotron      & Thermal dust        & Spinning dust \\
\hline \hline

Model 0b          & $r = 0.05$      &   N/A &  N/A    &   N/A       \\
\hline

Model 1a      & $r = 0.05$     &    Uniform & One MBB  &   N/A                   \\
\hline

Model 1b         & $r = 0.05$    &   Uniform & Two MBB     &   N/A                   \\
\hline

Model 1c          & $r = 0.05$     &    Non-uniform &  One MBB  & N/A                 \\
\hline

Model 1d          & $r = 0.05$     &    $+0.3$ curvature & One MBB  &  N/A                 \\
\hline

Model 1e          & $r = 0.05$     &    $-0.3$ curvature & One MBB  &  N/A                 \\
\hline

Model 1f     & $r = 0.05$      &    Uniform & One MBB  &  1\,\%                   \\
\hline
\hline

Model 0c       & $r = 0.001$  &    N/A  &  N/A   &   N/A          \\
\hline
\hline

Model 0a          & $r = 0$      &    N/A & N/A        &  N/A                  \\
\hline

Model 2a      & $r = 0$     &    Uniform & One MBB  &   N/A                   \\
\hline

Model 2b          &  $r = 0$    &    Uniform & Two MBB      &   N/A                \\
\hline

Model 2c          & $r = 0$     &    Non-uniform & One MBB  &  N/A                 \\
\hline

Model 2d       & $r = 0$   &  $+0.3$ curvature & One MBB &  N/A                 \\
\hline

Model 2e          & $r = 0$    &   $-0.3$ curvature & One MBB  &  N/A                \\
\hline

Model 2f   & $r = 0$     &   Uniform & One MBB  &  1\,\%                   \\
\hline
  \end{tabular}
  \caption{Summary of sky model simulations used in this work.}
  \label{tab:models}
\end{table}

\setlength{\tabcolsep}{2mm}  % put table column back to normal

\subsection{Set of simulations}

The set of sky models is summarized in Table~\ref{tab:models}. 
For example, Model 1b includes CMB B-modes with tensor-to-scalar ratio $r=0.05$, and is a nominal foreground polarization model: basically the thermal dust is parametrized by two modified blackbodies \citep{Finkbeiner1999}, and the synchrotron by a uniform power law. 

The nominal noise levels in each frequency channel for the variety of B-mode satellite concepts are listed in Table~\ref{tab:experiments} and co-added to the sky simulations. We assume that the noise is thermal (no $1/f$ noise) and uniformly distributed on the sky.

The alternative descriptions of the Galactic foreground polarization proposed in this set of simulations allows us to study the impact on the CMB B-modes of fitting an incorrect model for thermal dust, synchrotron, or spinning dust. Specifically, we test the impact of under-estimating the complexity of the thermal dust frequency dependence by fitting to the data a single MBB dust component whereas the sky consists of two MBB dust components, of neglecting the synchrotron curvature or the variability of the synchrotron spectral index, and of omitting low-polarized Galactic foregrounds such as the spinning dust radiation.

%%%%%%%%%%%%%%%%%%%%%%%%%%%%%%%%%%%%%%%%%%%%%%%%%%%%
%%%
%%% Method
%%%
\section{Method}
\label{sec:method}

\subsection{Separation of the component maps}
\label{subsec:compsep}

We perform the component separation and the reconstruction of the CMB map and power spectrum by implementing a Bayesian parametric fitting method using Gibbs sampling through the implementation of the {\tt Commander} algorithm \citep{Eriksen2008}.  

The model of the sky $\bdd(p,\nu)$, that we fit to the simulated data in direction $p$ (pixel) in the sky and at frequency $\nu$, is a linear combination of the CMB component, the thermal dust emission and the synchrotron emission: 
\bea
\label{eq:fit}
\bdd(p,\nu) &=& a(\nu)\,\bds^{cmb}(p) \cr
            &+& \left({\nu\over \nu_0^{s}}\right)^{\beta_s(p)}\, \bds^{sync}(p) \cr
            &+& \left({\nu\over\nu_0^{d}}\right)^{\beta_d(p)}{B_{\nu}\left(T_d(p)\right)\over B_{\nu_0^{d}}\left(T_d(p)\right)}\,\bds^{dust}(p) \cr
            &+& \bdn(p,\nu).
\eea 
We assume a uniform Gaussian instrumental noise $\bdn(p,\nu)$ for each experiment. Therefore, the fitting model is parametrized by
\bea
\bds = \left(\bds^{cmb},\bds^{dust},\bds^{sync}\right), \boldsymbol{\beta} = \left(\beta_d,T_d,\beta_s\right), C_\ell = \langle\vert\bds_{\ell m}^{cmb}\vert^2\rangle
\eea
where $\bds$ collects the $Q$, and $U$ maps of the CMB $\bds^{cmb}$, the thermal dust $\bds^{dust}$, and the synchrotron $\bds^{sync}$. The Galactic foregrounds are also parametrized by the spectral indices, $\beta_d$ for the thermal dust and $\beta_s$ for the synchrotron, and the temperature $T_d$ for the thermal dust. $C_\ell$ is the angular power spectrum of the CMB maps and can be written as
\bea
\label{eq:ps}
C_\ell =  \left(
\begin{array}{ccc}
C_\ell^{TT} & C_\ell^{TE} & C_\ell^{TB} \\
C_\ell^{TE} & C_\ell^{EE} & C_\ell^{EB}  \\
 C_\ell^{TB} & C_\ell^{EB} & C_\ell^{BB}  \\
\end{array}
\right).
\eea
While the CMB and the noise can be accurately parametrised from known theoretical models, Galactic foreground parametrisation can be subject to incorrect modelling due to imperfect knowledge of the foreground physics. In the present work, we play with the second and third lines of Eq. \ref{eq:fit} allowing for some mismatch between the fitted foreground model and the foreground data from the sky simulations for different B-mode experiments.

Using Bayes' theorem, the joint CMB-foreground posterior distribution is given by:
\bea
P\left(\bds,\boldsymbol{\beta},C_\ell \big| \bdd \right) \propto P\left(\bdd \big| \bds,\boldsymbol{\beta},C_\ell\right)P\left(\bds,\boldsymbol{\beta},C_\ell\right)
\eea 
where $P\left(\bds,\boldsymbol{\beta},C_\ell\right)$ is the prior distribution. The {\tt Commander} algorithm performs a Gibbs sampling on the Monte-Carlo Markov chains (MCMC) as
\bea
\label{eq:gibbs}
\bds^{(i+1)} &\leftarrow& P\left(\bds \big| C_\ell^{(i)},\boldsymbol{\beta}^{(i)},\bdd\right),\cr
C_\ell^{(i+1)} &\leftarrow& P\left(C_\ell \big| \bds^{(i+1)}\right),\cr
\boldsymbol{\beta}^{(i+1)} &\leftarrow& P\left(\boldsymbol{\beta} \big| \bds^{(i+1)}, \bdd\right),
\eea 
which converges to the sampling from the joint posterior $P\left(\bds,\boldsymbol{\beta},C_\ell \big| \bdd \right)$.

From the marginalized distribution of the CMB power spectrum 
\bea
P\left(C_\ell \big| \bdd\right) = \int P\left(\bds,\boldsymbol{\beta},C_\ell \big| \bdd \right) d\bds^{cmb} d\bds^{dust} d\bds^{sync} d\beta_d d\beta_s,
\eea
the mean CMB power spectrum 
\bea
\langle C_\ell \rangle\nonumber &= \int P\left(C_\ell \big| \bdd\right)C_\ell dC_\ell
\eea
therefore reduces to a simple ensemble average over the Gibbs samples:
\bea
\label{eq:cl}
\langle C_\ell \rangle &= {1\over N}\sum_{i=1}^N C_\ell^{(i)},
\eea
where $N$ is the number of samples in the Markov chain.
Similarly, the covariance matrix of the $C_\ell$ can be computed as the ensemble average over the Gibbs samples as
\bea
\label{eq:sigma}
\Sigma_{\ell} = {1\over N}\sum_{i=1}^N \left(C_\ell^i - \langle C_\ell \rangle\right)^2.
\eea

Since the sky is assumed Gaussian and isotropic, the conditional distribution used in Eq. \ref{eq:gibbs} to sample $C_\ell$ is an Inverse-Gamma distribution when interpreted as a function of $C_\ell$:
\bea
P\left(C_\ell \big| \bds\right) \propto {e^{-{\left(2\ell +1\right)\over 2C_\ell}\left({1\over 2l+1} \sum_{m=-\ell}^\ell \vert \bds_{\ell m} \vert^2\right)} \over C_\ell^{(2\ell+1)/2}},
\eea
for which a simple textbook algorithm exists \mbox{\citep{Eriksen2004}}.
The conditional distribution used to sample the amplitude of the sky components is a Gaussian distribution
\bea
P\left(\bds \big| C_\ell,\bdd\right) &\propto& P\left(\bdd \big| \bds,C_\ell\right)P\left(\bds \big| C_\ell\right)\cr
&\propto& e^{(-1/2)\left(\bdd - \bds\right)^T\bdN^{-1}\left(\bdd - \bds\right)}e^{(-1/2)\bds^T\bdS^{-1}\bds}\cr
&\propto& e^{(-1/2)\left(\bds - \widehat{\bds}\right)^T\left(\bdS^{-1}+\bdN^{-1}\right)\left(\bds - \widehat{\bds}\right)},
\eea
where $\bdS$ and $\bdN$ are the covariance matrices in pixel space of the CMB and the noise respectively, and $\widehat{\bds}$ is given by the Wiener filtered data
\bea
\widehat{\bds} = \left(\bdS^{-1}+\bdN^{-1}\right)^{-1}\bdN^{-1} \bdd.
\eea
The sample $\bds^{(i+1)}$ can therefore be derived from the sample $\bds^{(i)}$ as
the conjugate gradient solution of 
\bea\left(\bdS^{-1}+\bdN^{-1}\right)\bds = \bdN^{-1} \bdd + \bdS^{-1/2}w_0 + \bdN^{-1/2}w_1
\eea
where $w_0,w_1 \sim \mathcal N(0,1)$.

\subsection{Likelihood estimation of the cosmological parameters}
\label{subsec:lnl}

Once CMB power spectra $\widehat{C}_\ell^{EE}$ and $\widehat{C}_\ell^{BB}$ have been estimated from Eq. \ref{eq:cl}, as well as the covariance matrix $\widehat{\Sigma}_\ell$  from Eq. \ref{eq:sigma}, we can perform a likelihood analysis of the cosmological parameters $\tau$ (optical depth to Reionization) and $r$ (tensor-to-scalar ratio).

At low multipoles, e.g. $\ell \lesssim 12$, we make the following assumption on the CMB power spectrum:
\bea
\label{eq:approx}
C_{\ell~\lesssim~12}^{EE} \propto \tau^2,\quad C_{\ell~\lesssim~12}^{BB} \propto  r.
\eea
We then compute the likelihood distribution
\bea
\label{eq:like}
-2\ln \mathcal{L}\left( \widehat{C}_\ell \big| \tau, r\right) \propto \left[\widehat{C}_\ell - C_\ell^{th}\left(\tau,r\right)\right]\widehat{\Sigma}^{-1}\left[\widehat{C}_\ell - C_\ell^{th}\left(\tau,r\right)\right]
\eea
for different values of $\tau$ and $r$, i.e. for different amplitudes of a theoretical template of the CMB power spectrum, $C_\ell^{th}\left(\tau,r\right)$, (Eq \ref{eq:approx}). This allows us for drawing the likelihood distributions $P(\tau)$ and $P(r)$ respectively for the optical depth, $\tau$, and the tensor-to-scalar parameter, $r$. 

While $\tau$ can be constrained by the amplitude of E-modes, the tensor-to-scalar ratio $r$ relies on the sole amplitude of the CMB B-mode power spectrum, which unlike the amplitude of E-modes is many orders of magnitude smaller than the amplitude of the Galactic foreground polarization signal. For this reason $r$ is much more sensitive than $\tau$ to any imperfect foreground modelling. In the following, we show the impact of foregrounds on the tensor-to-scalar parameter only.

%%%%%%%%%%%%%%%%%%%%%%%%%%%%%%%%%%%%%%%%%%%%%%%%%%%%
%%%
%%% Results
%%%
\section{Results}
\label{sec:results}

In this section we compare the sensitivity to large-scale B-modes (tensor-to-scalar ratio) of different satellite concepts, COrE, COrE+, LiteBIRD, PIXIE, EPIC, and PRISM, first assuming no Galactic foreground, then considering Galactic foreground cleaning with the parametric fitting method described in Section \ref{subsec:compsep}. In particular, we show the impact on the estimated tensor-to-scalar ratio of incorrect assumptions on the Galactic foreground model. For all CMB experiments, we compute both the $\chi^2$ value from parametric fitting and the bias on the recovered tensor-to-scalar ratio resulting from incorrect foreground modelling.

We consider different mismatches between the foreground model and the simulated data: i) incorrect spectral modelling of the thermal dust through the assumption of a single MBB spectrum instead of two MBB spectra, ii) impact of high-frequency channels and Galactic masking, iii) incorrect spectral modelling of the synchrotron through the omission of a curvature term in the spectral distribution or through the incorrect assumption of a uniform spectral index, and iv) omission of an extra polarized component like spinning dust.

 The different sky simulations for polarized foregrounds, investigated in this work, are listed in Table~\ref{tab:models}. 
Regardless of the different sky simulations, we intentionally fit with {\tt Commander} a unique parametric foreground model to the data given by Eq \ref{eq:fit}: a single MBB component for the thermal dust, and a uniform power-law with zero curvature for the synchrotron. We adopt Gaussian priors for the dust spectral index $\beta_d = 1.6 \pm 0.3$, the dust temperature $T_d = 18 \pm 0.05$\,K, and the synchrotron spectral index $\beta_s = -3 \pm 0.1$. The Gaussian priors are multiplied by the Jeffreys prior to take into account the volume of likelihood space for non-linear parameters \citep{Eriksen2008}. Therefore, we are considering a possible mismatch between the model and the data regarding the spectral behaviour of the thermal dust and the synchrotron, and also regarding the omission of the spinning dust polarization.

The likelihood distributions of the tensor-to-scalar ratio for the different CMB satellite concepts and the different sky configurations considered in this work are plotted in Fig.~\ref{Fig:summary_r} (dust modelling),  in Fig.~\ref{Fig:summary_r2} (synchrotron modelling), and in Fig.~\ref{Fig:summary_r3} (spinning dust modelling). 
The maximum likelihood value of $r$ obtained for the different CMB satellite concepts and the different sky configurations, taken together with the $\chi^2$-values testing the quality of the Bayesian parametric fitting at the component separation stage, are listed in Tables~\ref{tab:results_dust}, \ref{tab:results_sync}, and \ref{tab:results_spin}, along with their $1\sigma$ errors.

\subsection{Overall sensitivity on large scales}
\label{subsec:sensitivity}

\subsubsection{No foregrounds, $r=0$}
\label{sec:no_foregrounds_r=0}

In the absence of any CMB B-modes ($r = 0$) and any Galactic foreground (model 14 in Table~\ref{tab:models}) the simulated observations of the sky at any frequency for any CMB experiment would be characterized by the combination of pure CMB E-mode polarization and instrumental noise. 

Such a simple configuration can provide forecasts on the instrumental sensitivity to B-modes of the different CMB satellite concepts considered in this work. In this case, the implementation of the {\tt Commander} algorithm reduces to a simple denoising of the CMB polarization in each frequency channel. The complete set of frequency channels is used in the Bayesian fitting: $15$ channels for COrE from $45$\,GHz to $795$\,GHz, $19$ channels for COrE+ Light from $60$\,GHz to $600$\,GHz, $21$ channels for COrE+ Extended from $60$\,GHz to $800$\,GHz, $6$ channels for LiteBIRD from $30$\,GHz to $280$\,GHz, $33$ channels for PIXIE from $30$\,GHz to $990$\,GHz (we do not consider channels beyond 1000\,GHz that are available to PIXIE since this would involve more complicated foreground modelling), $7$ channels for EPIC-CS from $30$\,GHz to $300$\,GHz, $8$ channels for EPIC-CS from $30$\,GHz to $500$\,GHz, $9$ channels for EPIC-IM-4K from $30$\,GHz to $850$\,GHz, $20$ channels for PRISM from $30$\,GHz to $660$\,GHz.   Depending on the different frequency coverages and the different instrumental sensitivities in each individual channel, the resulting overall sensitivity to B-modes may vary from one CMB satellite concept to another.

The r.m.s. of the B-mode polarization map resulting from pure instrumental noise residuals after component separation provides the overall sensitivity to B-modes for each CMB experiment. This exercise is interesting as a pure test of sensitivity because in the absence of CMB B-modes the error is no longer dominated by cosmic variance. The overall sensitivity to B-modes for all CMB satellite concepts is shown in the top left panel of Fig.~\ref{Fig:summary_r}. The experiments PRISM, EPIC-IM, and COrE+ Extended show the best instrumental sensitivity to B-modes while PIXIE and \mbox{EPIC-CS} show the largest r.m.s. noise in terms of B-mode polarization. It is interesting to note that, in the absence of any foreground and any lensing contamination, COrE and LiteBIRD have similar sensitivity to B-modes. This of course must be reconsidered in presence of foreground contamination (see hereafter) and/or lensing contamination; having more frequency channels than LiteBIRD, COrE is better designed than LiteBIRD to reduce uncertainties on $r$ after foreground cleaning. The reduced number of frequency channels from LiteBIRD may reduce its ability to clean contamination by non-trivial foregrounds. In addition, high-resolution experiments, like COrE, should be in a better position than low-resolution experiments, like LiteBIRD, to correct for lensing B-mode contamination and reduce the uncertainty on $r$ due to lensing.

\subsubsection{No foregrounds, $r=0.05$}

When the sky includes CMB B-mode polarization with a tensor-to-scalar ratio $r=0.05$ but still no Galactic foreground -- Model 0b in Table~\ref{tab:models} --  we do not need to fit for Galactic foregrounds at the component separation stage with {\tt Commander} but only for CMB and noise. The complete set of frequency channels of the CMB satellite concepts is used in the Bayesian fitting. The low-multipole CMB B-mode power spectrum resulting from component separation can be used afterwards to compute the likelihood distribution of the tensor-to-scalar ratio (Eqs \ref{eq:approx} and \ref{eq:like}) by using multipole modes up to the Reionization bump ($\ell \lesssim 12$).

As expected, in the absence of any Galactic foreground and a fortiori the absence of any assumptions about foregrounds, we get an unbiased estimate of the recovered tensor-to-scalar ratio (top right panel of Fig.~\ref{Fig:summary_r}).\footnote{Note that the template power spectrum used in the likelihood Eq. \ref{eq:like} is not the {\tt CAMB}  power spectrum with the true value $r=0.05$ but the power spectrum of the input CMB map realization of the simulation, for which the actual value of the realization is $r=0.042$. In all the plots of the paper, we deliberately rescale the realization value $r=0.042$ to the true value $r=0.05$ in order to highlight biases resulting from incorrect foregound modelling not from a particular CMB realization.}
In the absence of foregrounds, a $r = 0.05$ signal is detected at $8\sigma$ by PIXIE and at $9\sigma$ by all the other experiments.
When the sky includes CMB B-mode polarization with a tensor-to-scalar ratio $r=0.05$ but still no Galactic foregrounds, the error on the recovered tensor-to-scalar ratio is dominated by cosmic variance instead of instrumental noise for all CMB satellite concepts considered here. This is illustrated in the top right panel of Fig.~\ref{Fig:summary_r} where the likelihood distribution of the tensor-to-scalar ratio $P(r)$ show equal width for all CMB satellite concepts, despite the different instrumental characteristics of each satellite (also see Table~\ref{tab:results_dust} where the $1\sigma$ errors on $r$ have similar values for all CMB experiments). The sample variance can be reduced by analysing to higher $\ell$-values, however, this would then require de-lensing and can be achieved from ground- or balloon-based experiments.

\subsubsection{No foregrounds, $r=0.001$}

%%%%%%%%%%%%%%%%%%%%%%%%%%%%%%%%%%%%%%%%%%%%%%%%%%%%%%
\begin{figure*}
  \begin{center}
    \includegraphics[width=2\columnwidth]{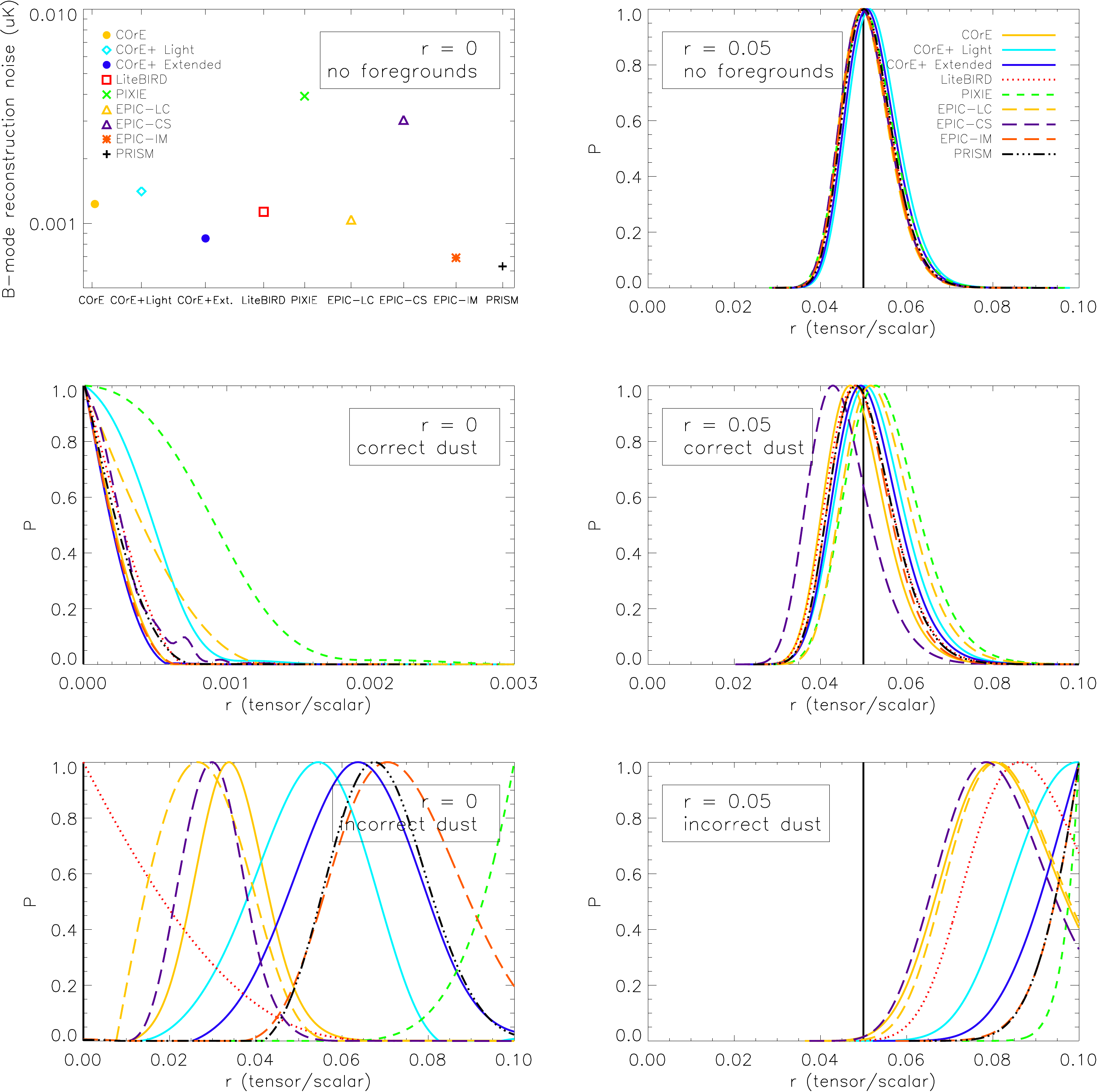}~
  \end{center}
\caption{Recovered posterior distribution $P(r)$ of the tensor-to-scalar ratio and impact of incorrect dust modelling. The theoretical input tensor-to-scalar value (vertical solid black line) is $r=0$ in the \emph{left-hand panels} and $r=0.05$ in the \emph{right-hand panels}. \emph{Top panels}: no foregrounds (\emph{left}: Model 0a, \emph{right}: Model 0b). \emph{Middle panels}: correct foreground modelling (\emph{left}: Model 2a, \emph{right}: Model 1a). \emph{Bottom panels}: incorrect spectral modelling of thermal dust (\emph{left}: Model 2b, \emph{right}: Model 1b). Recovered tensor-to-scalar distributions: COrE (solid yellow), COrE+ Light (solid light-blue), COrE+ Extended (solid blue), LiteBIRD (dotted red), PIXIE (dashed green), EPIC-LC-TES (long-dashed yellow), EPIC-CS (long-dashed purple), EPIC-IM-4K (long-dashed orange), PRISM (dash three-dot black). The \emph{top left panel} compares the overall sensitivity of the different satellites in the absence of foregrounds by showing for Model 0a ($r=0$, no foregrounds) the r.m.s of the residual noise B-mode map after component separation.}
\label{Fig:summary_r}
\end{figure*}
%%%%%%%%%%%%%%%%%%%%%%%%%%%%%%%%%%%%%%%%%%%%%%%%%%%%%%

Because of the high level of sensitivity of the CMB experiments considered here, even if we include in the likelihood analysis multipoles up to $\ell \lesssim 50$, the error on the estimated tensor-to-scalar ratio continues to be dominated by cosmic variance in the absence of any foreground, at least when $r=0.05$.

\begin{figure}
  \begin{center}
    \includegraphics[width=\columnwidth]{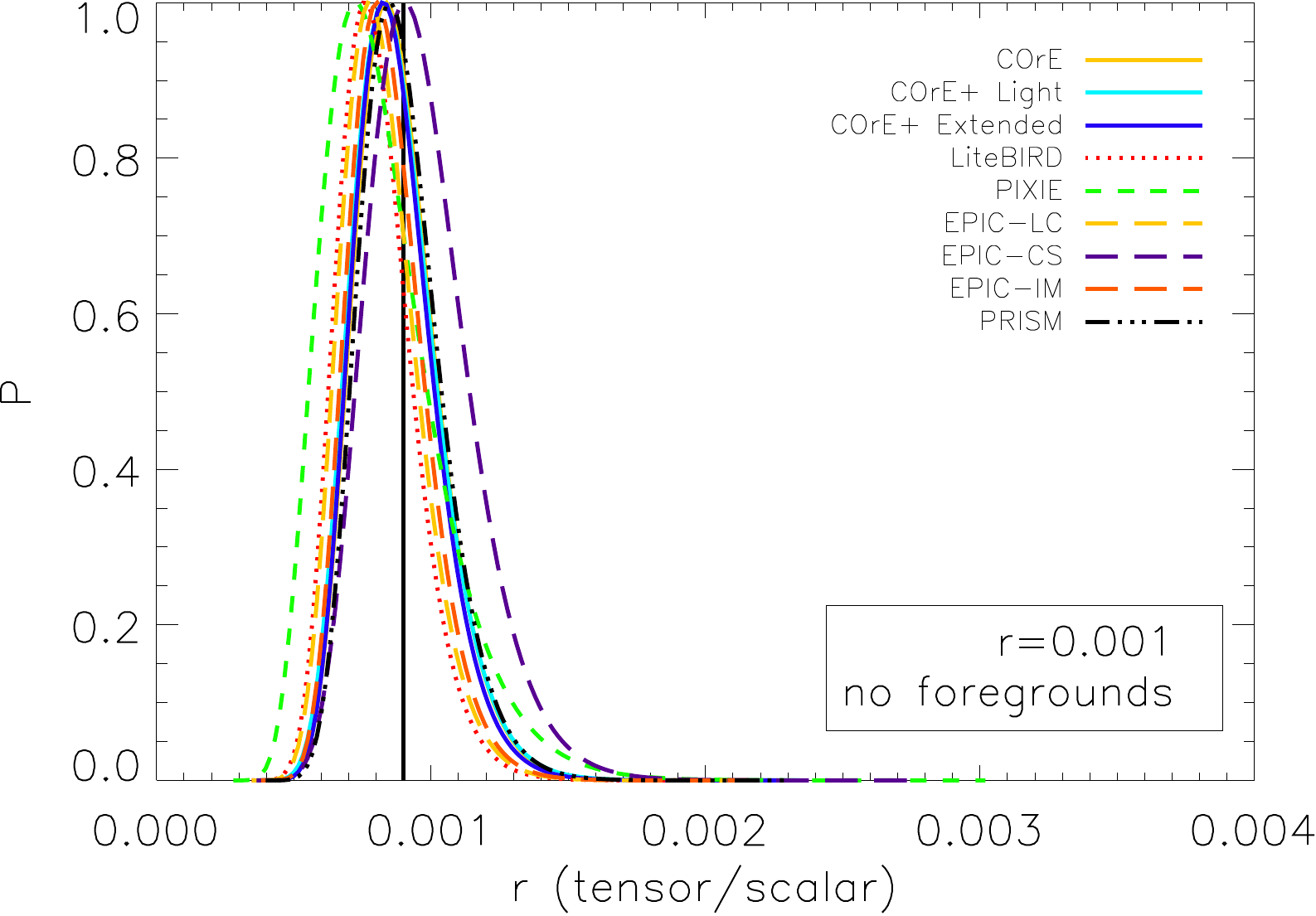}\\
 \end{center}
\caption{Recovered posterior distribution $P(r)$ of the tensor-to-scalar ratio for Model 0c (theoretical input $r=0.001$, no foregrounds): COrE (solid yellow), for COrE+ Light (solid light-blue), COrE+ Extended (solid blue), LiteBIRD (dotted red), PIXIE (dashed green), EPIC-LC-TES (long-dashed yellow), EPIC-CS (long-dashed purple), EPIC-IM-4K (long-dashed orange), and PRISM (dash three-dot black) experiments.}
\label{Fig:model29}
\end{figure}

In the case where the amplitude of the CMB B-modes would be smaller on large scales, say $r=0.001$, and there is still no foreground -- Model 0c -- the proportion in the error budget due to cosmic variance becomes mitigated with respect to the intrinsic noise of the experiments. In this case, each CMB experiment show distinct $1\sigma$ errors on the recovered tensor-to-scalar ratio (Fig.~\ref{Fig:model29} and Table~\ref{tab:model29}). In particular, the experiments PRISM, COrE+ Extended, and EPIC-IM, yield the smallest error on the tensor-to-scalar ratio estimate with a detection of a $r=0.001$ signal at $6\sigma$ in the absence of foreground/lensing, therefore confirming the sensitivity results presented in the top left panel of Fig.~\ref{Fig:summary_r}. 

These numbers should be considered with caution because in this forecast analysis we neglect any additional error caused by lensing B-modes. Especially when $r$ is low, the impact of lensing might dominate the error on $r$ over the
contribution from both instrumental noise and cosmic variance. Depending on the resolution of each experiment, the sensitivity to $r$ after delensing may vary from an experiment to another \citep{Smith2012}.

\subsubsection{Galactic foreground B-modes}

The unprecedented sensitivity of next-generation CMB satellite missions is required to detect the primordial CMB B-mode signal on large scales, in particular if the energy scale of Inflation is small, resulting in $r << 0.1$. As a consequence, the uncertainty in the estimation of the tensor-to-scalar ratio will no longer be dominated by instrumental noise on large scales. Rather, the main uncertainty on $r$ will come from the contamination by highly-polarized Galactic foregrounds.

Considering the presence of Galactic foreground polarization in the observations, we can successfully perform foreground cleaning / component separation by implementing the Bayesian parametric fitting method described in Section \ref{subsec:compsep}, as long as our fitting model correctly matches the real (simulated) data in terms of parametrization of Galactic foregrounds. 

Model 1a includes Galactic foreground polarization through synchrotron radiation with a uniform power-law spectral distribution ($\beta_s = -3$) and thermal dust radiation characterized by a single MBB emissivity ($\beta_d = 1.6$, $T_d=18$\,K), in addition to CMB polarization with $r=0.05$ and instrumental noise. In this case, we perform component separation by fitting a parametric model that perfectly matches the simulated sky, i.e. single MBB component for thermal dust and synchrotron power-law with uniform spectral index over the sky (Gaussian priors $\beta_s = -3\pm 0.1$, $\beta_d = 1.6\pm 0.3$, $T_d = 18\pm 0.05$\,K). Moreover, the complete set of frequency channels of each experiment is used for the fit. Model 2a is identical except that there are no CMB B-modes ($r=0$). 

The panels on the second row of Fig.~\ref{Fig:summary_r} show that, with no modelling error on the Galactic foreground parametrization (Models 1a and 2a), the tensor-to-scalar ratio is successfully recovered without any bias for all the CMB satellite experiments. By comparing the right panels of Fig.~\ref{Fig:summary_r} from the first row (Model 0b, no foreground) and the second row (Model 1a, including foregrounds), we can see that the presence of Galactic foregrounds inflates the error of the recovered tensor-to-scalar ratio. From the first two blocks of Table~\ref{tab:results_dust} (Models 0b and 1a) we can measure that the error inflation due to foreground contamination goes from $\approx 28$\,\% (i.e. the error is increased by a factor $1.28$) for PRISM to $\approx 40$\,\% (i.e. the error is increased by a factor $1.4$) for PIXIE. In presence of foregrounds, we find after component separation that $r=0.05$ is now detected at $6\sigma$ for PIXIE and EPIC-CS and detected at $6.5\sigma$ by LiteBIRD, COrE+ Light and EPIC-LC, while it is detected at $7\sigma$ by COrE, COrE+ Extended, EPIC-IM, and PRISM.

The left panel on the second row of Fig.~\ref{Fig:summary_r} (Model 2a, $r=0$) shows different widths for the likelihood distribution of the tensor-to-scalar ratio from each CMB experiment. This illustrates the ability of each experiment in removing the polarized foregrounds from the observations.  In particular, PIXIE shows less sensitivity than the other experiments because it is intrinsically less sensitive than the other experiments as shown in Sect.~\ref{sec:no_foregrounds_r=0}, with a $1\sigma$ error on $r$ of $\sigma_r=4.2 \times 10^{-4}$. Despite the large number of high-frequency channels of PIXIE, the uncertainty on $r$ is still surprisingly larger than for the other experiments. Even if we include in the fit the whole set of frequency channels of PIXIE ($>$ 1\,THz), yet the uncertainty on $r$ is not significantly reduced. Actually, we find that even if the thermal dust is best fitted with PIXIE channels, nevertheless in this particular configuration and simulation of the sky it is not the dust but the synchrotron contamination that dominates the increase in the uncertainty on $r$. Therefore, extra low-frequency channels would better help than high-frequency channels in reducing the uncertainty on $r$. In particular, by combining the sensitive low-frequency C-BASS data at $5$\,GHz \citep{Irfan2015} with PIXIE data we estimate that the error on $r=0.05$ would be reduced by a factor $1.4$, therefore compensating the increase of error due to synchrotron contamination.
COrE+ Extended shows the best overall sensitivity and ability in removing the foreground contamination with $\sigma_r=1.2 \times 10^{-4}$. The recovered values for $r$ and their uncertainties are given in the fifth block of Table~\ref{tab:results_dust}.

In principle, we can measure the quality of the Bayesian fit / component separation with the $\chi^2$ distribution:
\bea
\label{eq:chi2}
\chi^2 (p) = \sum_\nu \left({\bdd(p,\nu) - \bdm(p,\nu)\over \sigma_\nu(p)}\right)^2,
\eea
where the sum runs over frequency channels, $\sigma_\nu(p)$ represents noise in the data, and ${(\bdd(p,\nu) - \bdm(p,\nu))}$ is the mismatch between the observation data, $\bdd(p,\nu)$, and the parametric model, $\bdm(p,\nu)$, based on the spectral parameters. For a perfect fit, this quantity should be proportional to the number of degrees of freedom used in the fit, i.e. $d.o.f=(N_{ch} - N_{par})$ where $N_{ch}$ is the number of observation channels (twice the number of frequencies because of $Q$ and $U$ polarization observations) and $N_{par}$ the number of parameters of the model. However, the effective number of parameters is not clearly defined in this Bayesian framework because of the presence of priors constraining the parameter space \citep{Planck2013_XII}. For this reason, it is difficult to compute the exact reduced chi-square distribution, $\chi^2 / d.o.f$, for the goodness-of-fit. However, we can use in this work the $\chi^2$-values of Model 1a and Model 2a  (Table~\ref{tab:results_dust}) as the baseline $\chi^2$ for the perfect fit of the foregrounds. Similarly, the tensor-to-scalar ratio values of Model 1a and Model 2a (Table~\ref{tab:results_dust}) can be regarded as the baseline values indicating correct foreground modelling. 

We can see that the mean $\chi^2$ of the fit in the presence of Galactic foregrounds (Model 1a in Table~\ref{tab:results_dust}) have similar values that in the case with no foregrounds (Model 0b in Table~\ref{tab:results_dust}). Any departure from the baseline $\chi^2$-values of Model 1a and Model 2a indicates incorrect foreground modelling, which may result in a bias in the recovered tensor-to-scalar ratio. Conversely, if the recovered tensor-to-scalar ratio shows a bias but the $\chi^2$-value remains close to the baseline value, then the fitting is still good over the range of frequencies considered and the bias on the tensor-to-scalar ratio is due to a lack of some frequency channels for the experiment. In this case, more frequency channels / data points would be needed for fitting a more complex foreground model, taking into account more degrees of freedom (e.g. non-uniform spectral index, synchrotron curvature, etc.).

\subsection{Impact of incorrect foreground modelling}
\label{subsec:foreground}

As a counterpart of high sensitivity, next-generation CMB satellite experiments will be more sensitive to any incorrect assumptions on the Galactic foreground parameters in any component separation algorithm. In particular, incorrect foreground modelling may dramatically bias the estimation of tensor-to-scalar ratio. Thermal dust modelling might be critical in the high-frequency regime ($\gtrsim 100$\,GHz) of the observations, while synchrotron modelling and spinning dust polarization may play a non-negligible role at low frequency ($\lesssim 100$\,GHz) for the accurate estimation of the tensor-to-scalar ratio. 

%%%%%%%%%%%%%%%%%%%%%%%%%%%%%%%%%%%%%%%%%%%%%%%%%%%%%%
%%%%%%%%%%%%%%%%%%%%%%%%%%%%%%%%%%%%%%%%%%%%%%%%%%%%%%

%%%%%%%%%%%%%%%%%%%%%%%%%%%%%%%%%%%%%%%%%%%%%%%%%%%%%%
\begin{figure}
  \begin{center}
    \includegraphics[width=\columnwidth]{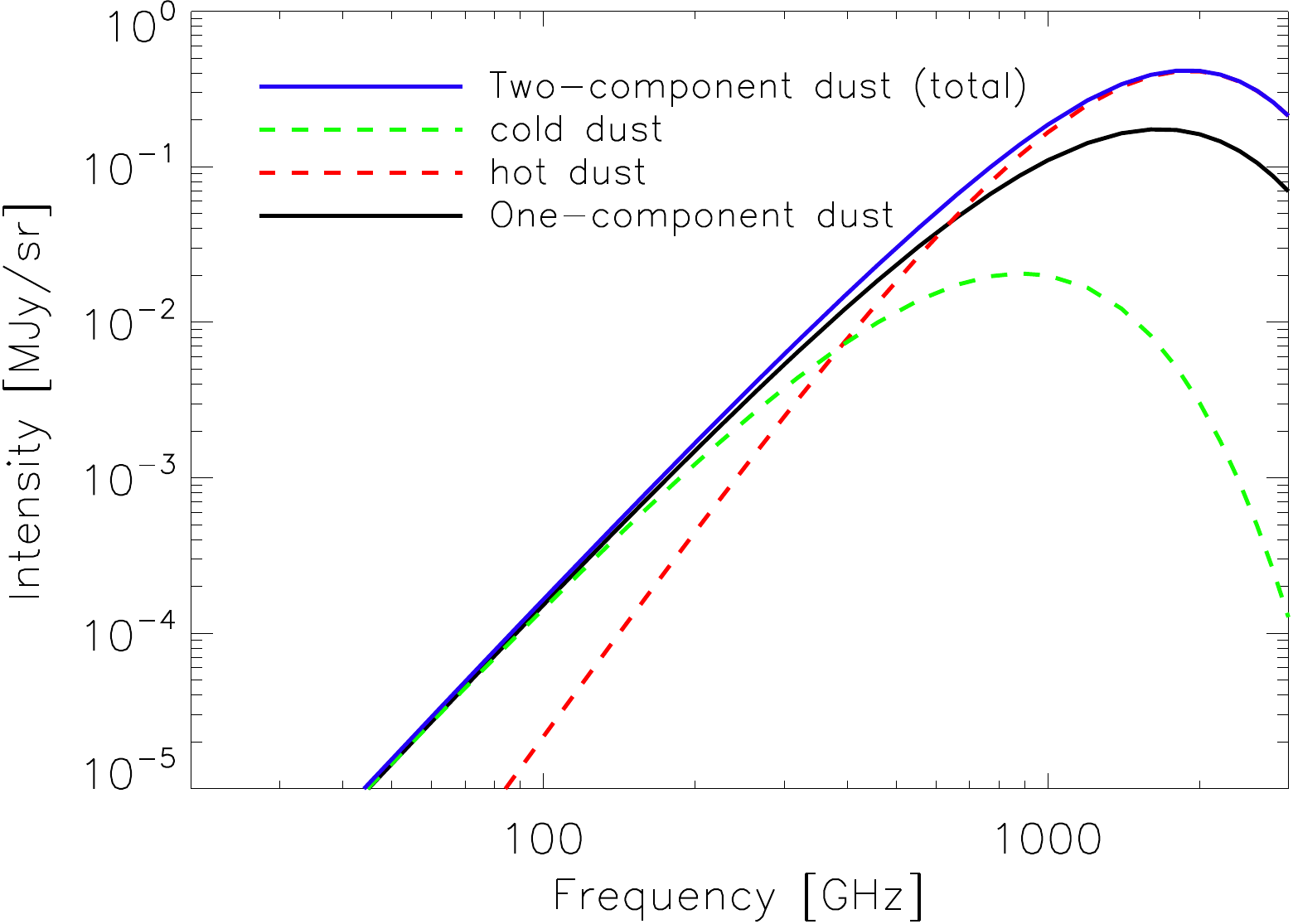}\\
 \end{center}
\caption{Modified blackbody (MBB) spectral energy distributions for thermal dust. The two-MBB dust model (solid blue) results from the coaddition of a cold dust component ($T=9.6$\,K, $\beta=1.5$ -- dashed green) and a hot dust component ($T=16.4$\,K, $\beta=2.6$ -- dashed red). 
The difference between the one-MBB dust model ($T=18$\,K, $\beta=1.6$ -- solid black), adopted in the Bayesian fitting, and the two-MBB dust model (solid blue), adopted in Model 1b and Model 2b sky simulations, is becoming more significant at frequencies larger than $300$\, GHz.}
\label{Fig:mbb}
\end{figure}
%%%%%%%%%%%%%%%%%%%%%%%%%%%%%%%%%%%%%%%%%%%%%%%%%%%%%%

\setlength{\tabcolsep}{2mm}

\begin{table*}
\centering
\begin{tabular}{l|l|l|l|l}
\hline
Model ID & Mean $\chi^2$ & Mean $\chi^2 / N_{ch}$ & Recovered $r$ & Experiment\\
\hline \hline
Model 0b & 37.51 & 0.99 &  0.05271 $\pm$ 0.00595 & COrE+ Light \\
 & 41.50 & 0.99 &  0.05202 $\pm$ 0.00585 & COrE+ Extended \\
~ $r=0.05$  & 29.48 & 0.98 &  0.05107 $\pm$ 0.00575 & COrE \\
~ no foreground & 11.53 & 0.96 &  0.05132 $\pm$ 0.00578 & LiteBIRD \\
& 65.51 & 0.99 &  0.05145 $\pm$ 0.00616 & PIXIE \\
 & 13.56 & 0.97 &  0.05074 $\pm$ 0.00572 & EPIC-LC-TES \\
 & 15.44 & 0.96 &  0.05086 $\pm$ 0.00583 & EPIC-CS \\
 & 17.44 & 0.97 &  0.05096 $\pm$ 0.00572 & EPIC-IM-4K \\
 & 39.58 & 0.99 &  0.05140 $\pm$ 0.00578 & PRISM \\
\hline
Model 1a & 37.87 & 1.00 &  0.05224 $\pm$ 0.00800 & COrE+ Light \\
 & 41.71 & 0.99 &  0.05132 $\pm$ 0.00770 & COrE+ Extended \\
~ $r=0.05$  & 29.72 & 0.99 &  0.04877 $\pm$ 0.00715 & COrE \\
~ correct foreground & 11.86 & 0.99 &  0.04997 $\pm$ 0.00767 & LiteBIRD \\
~ modelling & 65.67 & 1.00 &  0.05507 $\pm$ 0.00865 & PIXIE \\
 & 14.82 & 1.06 &  0.05411 $\pm$ 0.00823 & EPIC-LC-TES \\
 & 15.57 & 0.97 &  0.04506 $\pm$ 0.00730 & EPIC-CS \\
 & 17.65 & 0.98 &  0.04989 $\pm$ 0.00708 & EPIC-IM-4K \\
 & 39.82 & 1.00 &  0.05027 $\pm$ 0.00737 & PRISM \\
\hline
Model 1b & 32.14 & 1.07 &  0.08929 $\pm$ 0.00766 & COrE+ Light \\
 & 35.90 & 1.20 &  0.09218 $\pm$ 0.00624 & COrE+ Extended \\
~ $r=0.05$  & 22.32 & 1.12 &  0.08023 $\pm$ 0.01045 & COrE \\
~ incorrect dust & 13.19 & 1.10 &  0.08428 $\pm$ 0.00935 & LiteBIRD \\
~ modelling & 25.86 & 1.08 &  0.09711 $\pm$ 0.00265 & PIXIE \\
 & 18.47 & 1.32 &  0.08113 $\pm$ 0.00999 & EPIC-LC-TES \\
 & 15.08 & 1.08 &  0.07911 $\pm$ 0.01048 & EPIC-CS \\
 & 40.30 & 2.88 &  0.09434 $\pm$ 0.00485 & EPIC-IM-4K \\
 & 50.44 & 1.58 &  0.09446 $\pm$ 0.00467 & PRISM \\
\hline
\hline
Model 0a & 37.49 & 0.99 &  0.00000 $\pm$ 0.00005 & COrE+ Light \\
 & 41.49 & 0.99 &  0.00000 $\pm$ 0.00002 & COrE+ Extended \\
~ $r=0$  & 29.47 & 0.98 &  0.00000 $\pm$ 0.00004 & COrE \\
~ no foreground & 11.52 & 0.96 &  0.00000 $\pm$ 0.00003 & LiteBIRD \\
& 65.57 & 0.99 &  0.00000 $\pm$ 0.00036 & PIXIE \\
 & 13.55 & 0.97 & 0.00000 $\pm$ 0.00003 & EPIC-LC-TES \\
 & 15.43 & 0.96 & 0.00000 $\pm$ 0.00022 & EPIC-CS \\
 & 17.44 & 0.97 & 0.00000 $\pm$ 0.00001 & EPIC-IM-4K \\
 & 39.57 & 0.99 & 0.00000 $\pm$ 0.00001 & PRISM \\
\hline
Model 2a & 37.83 & 1.00 &  0.00031 $\pm$ 0.00024 & COrE+ Light \\
 & 41.66 & 0.99 &  0.00016 $\pm$ 0.00012 & COrE+ Extended \\
~ $r=0$  & 29.72 & 0.99 &  0.00017 $\pm$ 0.00013 & COrE \\
~ correct foreground & 11.85 & 0.99 &  0.00020 $\pm$ 0.00017 & LiteBIRD \\
~ modelling & 65.66 & 0.99 &  0.00056 $\pm$ 0.00042 & PIXIE \\
 & 14.88 & 1.06 &  0.00033 $\pm$ 0.00026 & EPIC-LC-TES \\
 & 15.57 & 0.97 &  0.00023 $\pm$ 0.00020 & EPIC-CS \\
 & 17.60 & 0.98 &  0.00016 $\pm$ 0.00013 & EPIC-IM-4K \\
 & 39.78 & 0.99 &  0.00019 $\pm$ 0.00015 & PRISM \\
\hline
Model 2b & 32.19 & 1.07 &  0.05229 $\pm$ 0.01223 & COrE+ Light \\
 & 35.95 & 1.20 &  0.06357 $\pm$ 0.01332 & COrE+ Extended \\
~ $r=0$  & 22.42 & 1.12 &  0.03453 $\pm$ 0.00821 & COrE \\
~ incorrect dust & 13.23 & 1.10 &  0.01595 $\pm$ 0.01249 & LiteBIRD \\
~ modelling & 25.87 & 1.08 &  0.09246 $\pm$ 0.00635 & PIXIE \\
 & 18.55 & 1.33 &  0.02792 $\pm$ 0.00936 & EPIC-LC-TES \\
 & 15.16 & 1.08 &  0.03018 $\pm$ 0.00699 & EPIC-CS \\
 & 40.38 & 2.88 &  0.07251 $\pm$ 0.01265 & EPIC-IM-4K \\
 & 50.52 & 1.58 &  0.06885 $\pm$ 0.01068 & PRISM \\
\hline
\end{tabular}
\caption{Maximum likelihood estimate of the tensor-to-scalar ratio $r$ (fourth column) for different experiments for sky simulations with baseline $r=0.05$ (resp. $r=0$) and $\tau = 0.087$. Model 0b: no foregrounds, $r=0.05$. Model 1a: correct foregrounds, $r=0.05$. Model 1b: incorrect thermal dust, $r=0.05$. Model 0a: no foregrounds, $r=0$ (no CMB B-modes).  Model 2a: correct foregrounds, $r=0$ (no CMB B-modes). Model 2b: incorrect thermal dust, $r=0$ (no CMB B-modes). In addition, mean $\chi^2$ values (second column) and normalized $\chi^2$ values (third column) from Bayesian parametric fitting for the separation of the component maps. $N_{ch}$ denotes the number of channels for each experiment (twice the number of frequencies because of $Q$ and $U$ polarization maps).}
\label{tab:results_dust}
\end{table*}

\setlength{\tabcolsep}{2mm}

\begin{table*}
\centering
\begin{tabular}{l|l|l|l|l}
\hline
Model ID & Mean $\chi^2$ & Mean $\chi^2 / N_{ch}$ & Recovered $r$ & Experiment\\
\hline \hline
Model 0c & 37.57 & 0.99 &  0.00104 $\pm$ 0.00019 & COrE+ Light \\
 & 41.57 & 0.99 &  0.00103 $\pm$ 0.00018 & COrE+ Extended \\
~ $r=0.001$  & 29.48 & 0.98 &  0.00104 $\pm$ 0.00019 & COrE \\
~ no foreground & 11.56 & 0.96 &  0.00095 $\pm$ 0.00017 & LiteBIRD \\
& 65.56 & 0.99 &  0.00098 $\pm$ 0.00026 & PIXIE \\
 & 13.56 & 0.97 &  0.00097 $\pm$ 0.00017 & EPIC-LC-TES \\
 & 15.43 & 0.96 &  0.00114 $\pm$ 0.00023 & EPIC-CS \\
 & 17.45 & 0.97 &  0.00100 $\pm$ 0.00017 & EPIC-IM-4K \\
 & 39.59 & 0.99 &  0.00106 $\pm$ 0.00018 & PRISM \\
\hline
\end{tabular}
\caption{Maximum likelihood estimate of the tensor-to-scalar ratio $r$ (fourth column) for different experiments for a sky simulation with baseline $r=0.001$ and no foregrounds (Model 0c). In addition, mean $\chi^2$ values (second column) and normalized $\chi^2$ values (third column) from Bayesian parametric fitting for the separation of the component maps. $N_{ch}$ denotes the number of channels for each experiment (twice the number of frequencies because of $Q$ and $U$ polarization maps).}
\label{tab:model29}
\end{table*}

\subsubsection{Incorrect spectral modelling of thermal dust}
\label{subsubsec:dust}

We now consider a configuration of the sky -- Model 1b and Model 2b -- where the thermal dust polarization consists in two MBB spectral components \citep{Finkbeiner1999}: a cold dust component with temperature $T_1 = 9.6$\,K and spectral index $\beta_1 = 1.5$, and a hot dust component with temperature $T_2 = 16.4$\,K and spectral index $\beta_2 = 2.6$. The synchrotron emission still has a uniform power-law distribution over frequencies, and the CMB either include B-mode polarization ($r=0.05$ for Model 1b) or not ($r=0$ for Model 2b).
Therefore, we are considering spectral mismatch between the simulated thermal dust (two MBB components) and the dust model Eq. \ref{eq:fit} (one MBB dust component) fitted to the data. 

In order to minimize the impact of incorrect dust modelling, we exclude from the parametric fitting any frequency channel above $360$\,GHz for all the CMB experiments. This is because at higher frequencies the effects of dust temperature variations have a significant effect on the spectrum. We therefore restrict the analysis to $10$ channels for COrE from $45$\,GHz to $315$\,GHz, $15$ channels for COrE+ Light from $60$\,GHz to $340$\,GHz, $15$ channels for COrE+ Extended from $60$\,GHz to $340$\,GHz, $6$ channels for LiteBIRD from $30$\,GHz to $280$\,GHz, and $12$ channels for PIXIE from $60$\,GHz to $360$\,GHz, $7$ channels for EPIC-CS from $30$\,GHz to $300$\,GHz, $7$ channels for EPIC-CS from $30$\,GHz to $340$\,GHz, $7$ channels for EPIC-IM-4K from $30$\,GHz to $340$\,GHz, $16$ channels for PRISM from $30$\,GHz to $320$\,GHz.   

%%%%%%%%%%%%%%%%%%%%%%%%%%%%%%%%%%%%%%%%%%%%%%%%%%%%%%
\begin{figure*}
  \begin{center}
    \includegraphics[width=\columnwidth]{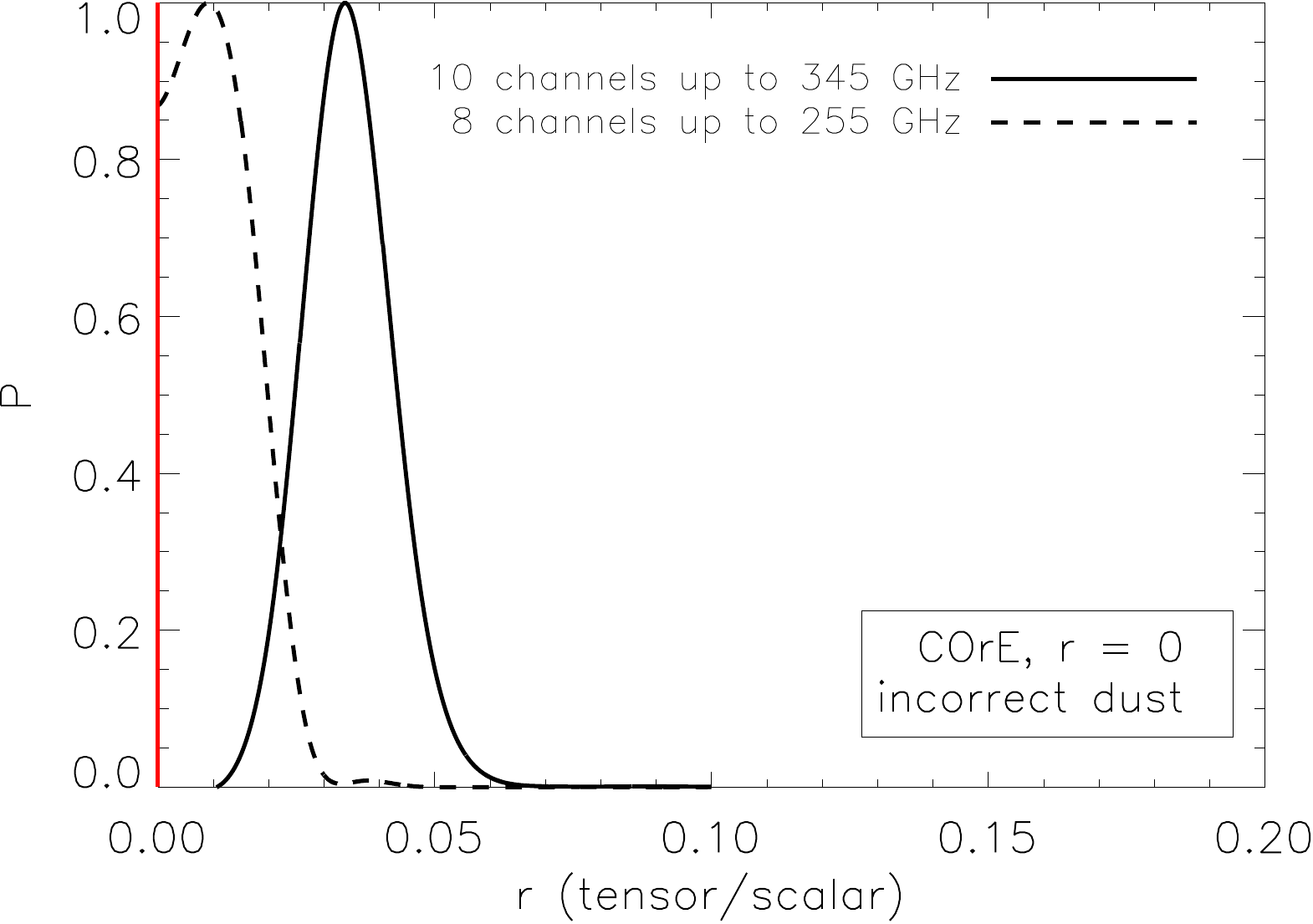}~
    \includegraphics[width=\columnwidth]{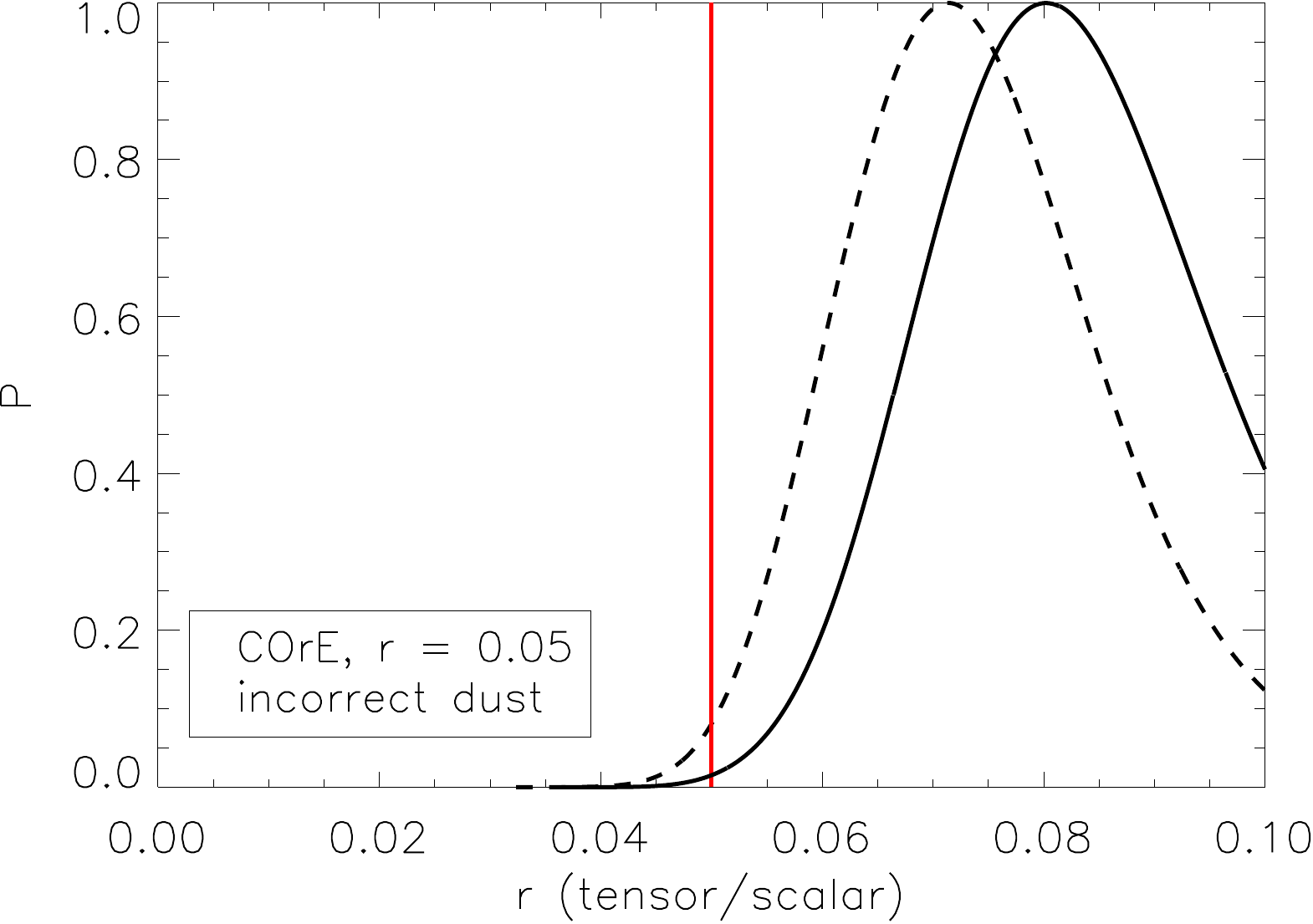}\\
 \end{center}
\caption{Impact of high-frequency channels on the estimation of the tensor-to-scalar ratio $r$ with COrE when incorrectly modelling the polarized thermal dust emission. The {\it solid black line} is for the frequency range 45--255\,GHz and the {\it dashed line} is when extending the frequency range to 345\,GHz. \emph{Left panel}: Model 2b ($r = 0$, incorrect dust spectrum). \emph{Right panel}: Model 1b ($r = 0.05$, incorrect dust spectrum). The high-frequency channels of COrE are useful to highlight any failure of the dust model. The instability of the mean value of the tensor-to-scalar distribution when including or not high-frequency channels provides a good diagnosis for revealing spurious foreground B-mode detection due to incorrect modelling.}
\label{Fig:bands}
\end{figure*}
%%%%%%%%%%%%%%%%%%%%%%%%%%%%%%%%%%%%%%%%%%%%%%%%%%%%%%

The results for the recovered distribution of the tensor-to-scalar ratio $P(r)$ are shown in the third row of Fig.~\ref{Fig:summary_r} for $r=0$ (left-hand panel) and $r=0.05$ (right-hand panel). 
Because of incorrect spectral assumptions on the number of thermal dust components in the parametric model, the recovered tensor-to-scalar ratio is found to be strongly biased, by more than 3$\sigma$ for all the experiments. The impact is even more significant for the experiments having the largest sensitivity at high frequencies, i.e. PRISM, EPIC-IM, and COrE+ Extended missions: the recovered tensor-to-scalar ratio is in this case biased by more than 5$\sigma$ when incorrectly parametrizing the spectral properties of the thermal dust. The maximum likelihood estimates of $r$ with 1$\sigma$ errors are listed in this case in the third (Model 1b) and sixth (Model 2b) blocks of Table~\ref{tab:results_dust}, showing strong departure from the baseline value of the tensor-to-scalar ratio. The corresponding mean $\chi^2$ values of the parametric fitting are listed in the third and sixth blocks of Table~\ref{tab:results_dust}: they are significantly larger than the baseline $\chi^2$ values of Model 1a (second block) and model 24 (fifth block), for which the foreground model was consistent with the data. This indicates a failure in the parametrization of the foreground model fitted to the data, and an unsuccessful separation of CMB and dust B-modes.

Clearly, the over-prediction of the tensor-to-scalar ratio is due to contamination by spurious dust foreground B-modes. The amount of spurious dust B-modes detected potentially increases for experiments having more high-frequency channels and more sensitivity because spectral modelling of thermal dust becomes more and more critical. Being limited to $255$\,GHz, LiteBIRD is less impacted by incorrect dust modelling, at least when $r=0$ (left-hand panel of the third row of Fig.~\ref{Fig:summary_r}), because at lower frequencies the coadded spectral distribution of the two MBB dust components is still consistent with the single MBB distribution of the model, as illustrated in Fig.~\ref{Fig:mbb}.  However, the increased $\chi^2=1.10$ value of LiteBIRD (Model 2b in Table~\ref{tab:results_dust}) with respect to the baseline $\chi^2=0.99$ value (Model 2a in Table~\ref{tab:results_dust}) still reveals incorrect assumptions on thermal dust, in which case the estimate of $r$ can not be considered as robust.
For PIXIE the bias on $r=0$ increases dramatically by more than $14\sigma$ (Model 2b in Table~\ref{tab:results_dust}) while the $\chi^2=1.08$ value stays close to the baseline value because of the larger noise level in the PIXIE channels compared to other experiments. This result shows that low sensitivity in a large number of channels may prevent an experiment from detecting incorrect foreground modelling, therefore leading to a false detection of $r$. However, in the case of PIXIE, the low sensitivity per channel can be compensated by including more frequency channels. 
In summary, experiments with more high/low-frequency channels and more sensitivity in high/low-frequencies, like PRISM, COrE, COrE+ are better designed to control foreground uncertainties and guarantee a robust detection of B-modes.

It is interesting to compare these results with those obtained on {\it Planck} simulations by \citet{Armitage-Caplan2012}: modelling a two-component dust simulation with a one-component dust model had only a minor impact on {\it Planck}, biasing the tensor-to-scalar ratio by less than 1$\sigma$ because of relatively lower sensitivity. It is clear that because next-generation CMB satellite missions (e.g. COrE, COrE+, LiteBIRD, PIXIE, EPIC, PRISM) will have significantly more sensitivity, they will be much more sensitive to any incorrect spectral modelling of thermal dust, as demonstrated in the third row of Fig.~\ref{Fig:summary_r}.

%%%%%%%%%%%%%%%%%%%%%%%%%%%%%%%%%%%%%%%%%%%%%%%%%%%%%%
\begin{figure}
  \begin{center}
    \includegraphics[width=\columnwidth]{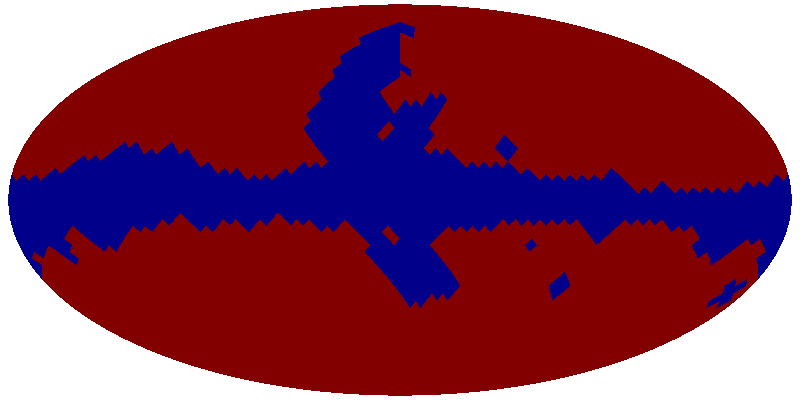}~\\
    \includegraphics[width=\columnwidth]{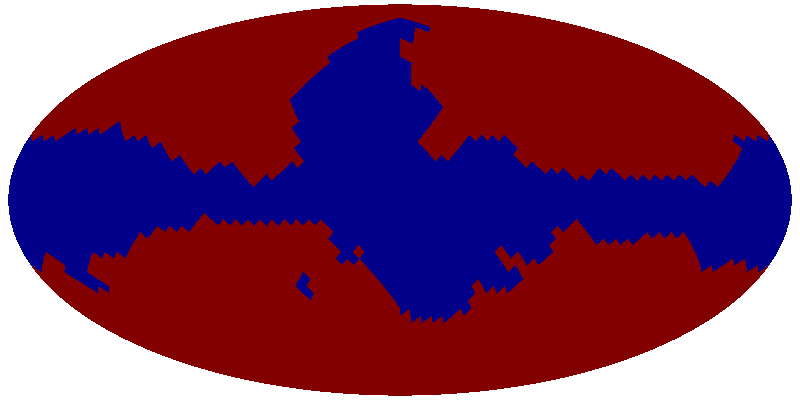}\\
 \end{center}
\caption{\emph{Top panel}: {\it WMAP} Low Resolution Polarization Data Analysis Mask \citep{Bennett2013} with $f_{\rm sky} = 73$\,\%. \emph{Bottom panel}: {\it Planck}-based polarization mask with $f_{\rm sky} = 63$\,\%. Blue regions are masked.}
\label{Fig:mask_maps} 
\end{figure}
%%%%%%%%%%%%%%%%%%%%%%%%%%%%%%%%%%%%%%%%%%%%%%%%%%%%%%

%%%%%%%%%%%%%%%%%%%%%%%%%%%%%%%%%%%%%%%%%%%%%%%%%%%%%%
\begin{figure*}
  \begin{center}
    \includegraphics[width=\columnwidth]{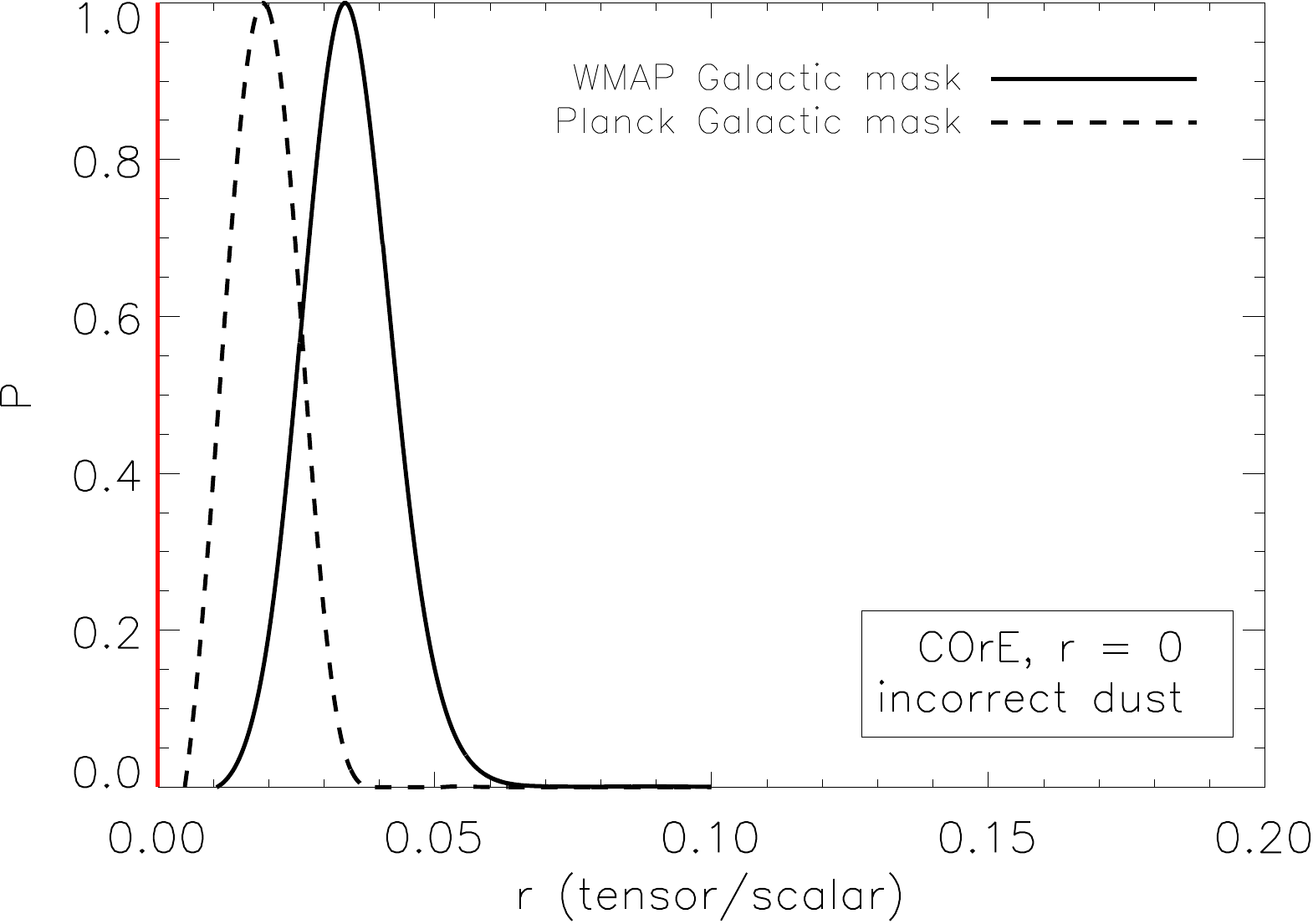}~
    \includegraphics[width=\columnwidth]{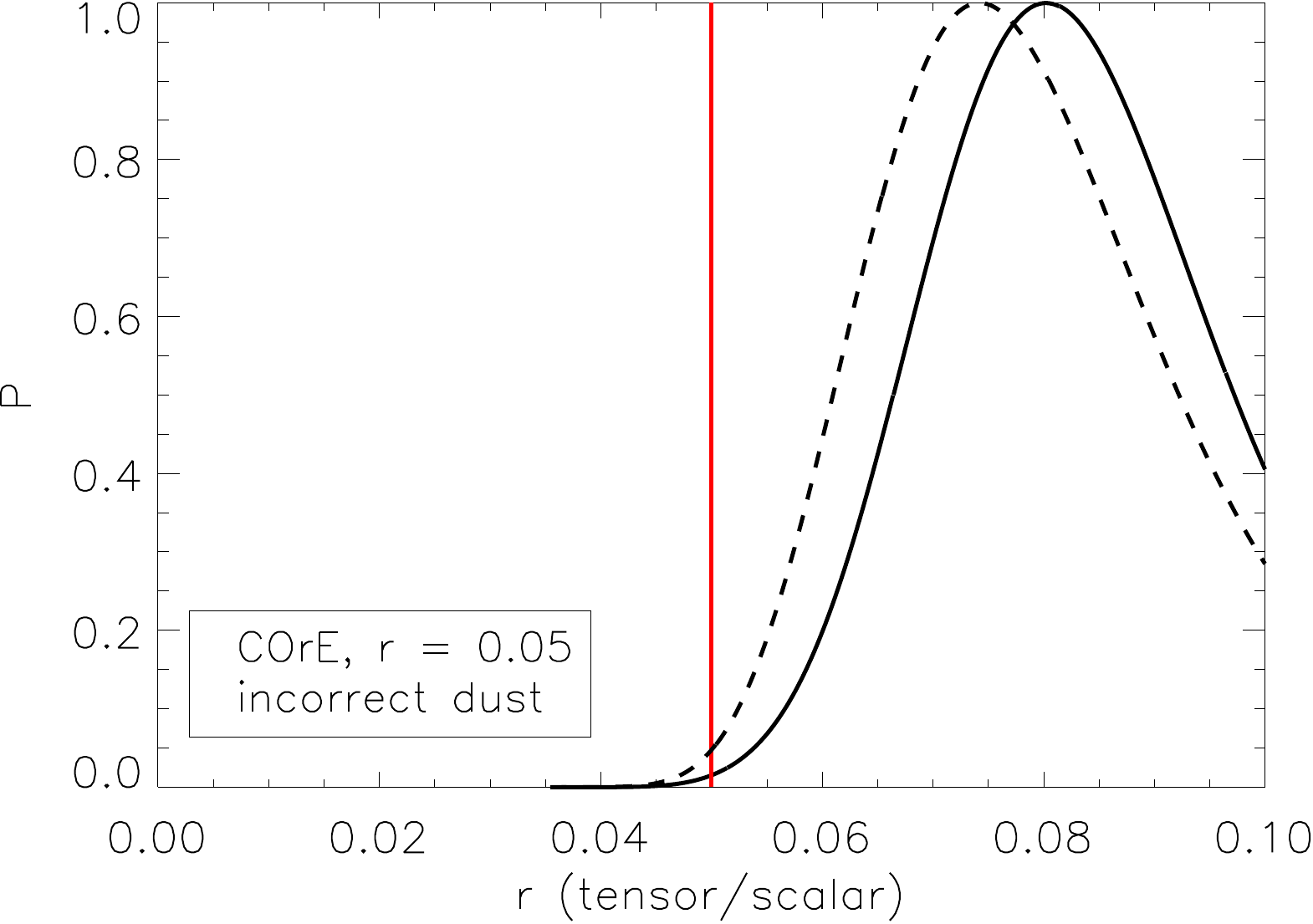}\\
 \end{center}
\caption{Impact of Galactic masking on the estimation of the tensor-to-scalar ratio $r$ with COrE when incorrect modelling the thermal dust spectrum. The {\it solid black line} is for the {\it Planck} mask (63\,\% coverage) and the {\it black dashed line} is for the WMAP mask (73\,\% coverage). \emph{Left panel}: Model 2b ($r = 0$, incorrect dust model). \emph{Right panel}: Model 1b ($r = 0.05$, incorrect dust model). The {\it Planck} mask, designed for masking the thermal dust intensity distribution over sky, reduces the amount of bias in the estimated tensor-to-scalar ratio.}
\label{Fig:masks} 
\end{figure*}
%%%%%%%%%%%%%%%%%%%%%%%%%%%%%%%%%%%%%%%%%%%%%%%%%%%%%%

The impact on the tensor-to-scalar ratio of incorrect dust modelling is more significant for CMB satellites having high-frequency observations. This is due to the increasing significance of the polarized dust emission at high frequency, where spectral parametrization becomes critical for this foreground. From the results shown on the third row of Fig.~\ref{Fig:summary_r}, it is clear that high-frequency channel observations ($350$ to $850$\,GHz) are particularly useful for CMB satellite missions to highlight any failure in the modelling of the polarized thermal dust emission.

%%%%%%%%%%%%%%%%%%%%%%%%%%%%%%%%%%%%%%%%%%%%%%%%%%%%%%
\begin{figure*}
  \begin{center}
    \includegraphics[width=2\columnwidth]{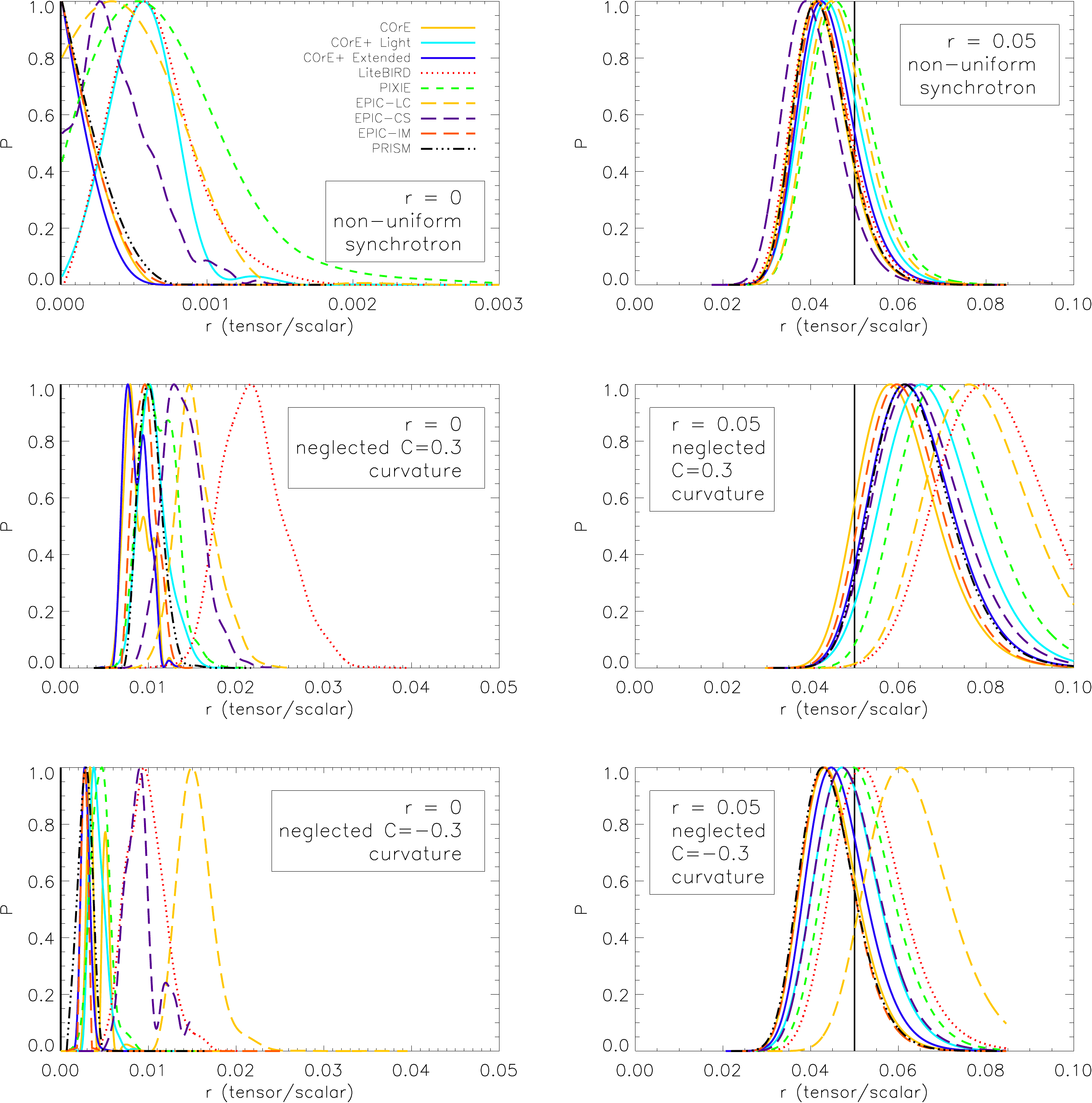}~
  \end{center}
\caption{Recovered posterior distribution $P(r)$ of the tensor-to-scalar ratio and impact of incorrect synchrotron modelling. The theoretical input tensor-to-scalar value (vertical solid black line) is $r=0$ in the \emph{left-hand panels} and $r=0.05$ in the \emph{right-hand panels}. \emph{Top panels}: neglected variability of the synchrotron spectral index (\emph{left}: Model 2c, \emph{right}: Model 1c). \emph{Middle panels}: neglected synchrotron curvature $C=+0.3$ (\emph{left}: Model 2d, \emph{right}: Model 1d). \emph{Bottom panels}: neglected synchrotron curvature $C=-0.3$ (\emph{left}: Model 2e, \emph{right}: Model 1e). Recovered tensor-to-scalar distributions: COrE (solid yellow), COrE+ Light (solid light-blue), COrE+ Extended (solid blue), LiteBIRD (dotted red), PIXIE (dashed green), EPIC-LC-TES (long-dashed yellow), EPIC-CS (long-dashed purple), EPIC-IM-4K (long-dashed orange), PRISM (dash three-dot black).}
\label{Fig:summary_r2}
\end{figure*}
%%%%%%%%%%%%%%%%%%%%%%%%%%%%%%%%%%%%%%%%%%%%%%%%%%%%%%

Taking the COrE satellite as an example, Fig.~\ref{Fig:bands} shows the effect of frequency coverage on the bias on $r$ when incorrectly parametrizing the thermal dust foreground. The amount of bias in the recovered tensor-to-scalar ratio with respect to the original value $r=0$ (left-hand panel) or $r=0.05$ (right-hand panel) increases when including in the parametric fitting frequencies larger than $255$\,GHz. 

The check of instability of the mean value of the tensor-to-scalar distribution $P(r)$ when including or not higher frequency channels in the Bayesian analysis, as shown in Fig.~\ref{Fig:bands}, provides a robust test of spurious foreground B-mode detection and incorrect foreground modelling.

The amount of bias of the recovered tensor-to-scalar ratio due to incorrect foreground modelling is also dependent on Galactic masking. To illustrate this, we use two different Galactic masks (Fig.~\ref{Fig:mask_maps}): the {\it WMAP} Low Resolution Polarization Data Analysis Mask\footnote{\url{http://lambda.gsfc.nasa.gov/product/map/dr5/masks_get.cfm}} \citep{Bennett2013}, relying mainly on synchrotron intensity thresholds over the sky from the {\it WMAP}\, K band polarized intensity map, and a more conservative polarization mask, based on both synchrotron and thermal dust intensity thresholds from the {\it Planck} polarized intensity maps at $30$\,GHz and $353$\,GHz \citep{Planck2015_X}. The synchrotron intensity threshold is determined as follows. We simulate CMB Q and U polarization maps at $70$\,GHz and $5$ degree beam resolution and we compute the RMS of the polarized intensity of the CMB, $\sigma(\sqrt{Q^2+U^2})=0.4\,{\rm\mu K}$, which is taken as a reference. We smooth the {\it Planck} $30$\,GHz polarization intensity map at $5$ degree resolution and extrapolate it to $70$\,GHz through a power-law with spectral index $\beta=-3$. We compute its RMS and mask any pixel where the RMS of the {\it Planck} $30$\,GHz polarization intensity extrapolated to $70$\,GHz is 2.5 larger or more than the reference RMS of the CMB polarization intensity. The dust intensity threshold is determined in the exact same way by using the {\it Planck} $353$\,GHz polarization map and extrapolating it throuh a power-law with spectral index $\beta=1.6$. The {\it Planck}-based conservative mask is defined by the union of the synchrotron mask and the dust mask.
The fraction of the observed sky is $63$\,\% for the {\it Planck}-based conservative mask and $73$\,\% for the {\it WMAP} mask.

\setlength{\tabcolsep}{2mm}

\begin{table*}
\centering
\begin{tabular}{l|l|l|l|l}
\hline
Model ID & Mean $\chi^2$ & Mean $\chi^2 / N_{ch}$ & Recovered $r$ & Experiment\\
\hline \hline
Model 1c & 37.88 & 1.00 &  0.05343 $\pm$ 0.00830 & COrE+ Light \\
 & 41.73 & 0.99 &  0.05212 $\pm$ 0.00777 & COrE+ Extended \\
~ $r=0.05$  & 29.78 & 0.99 &  0.05049 $\pm$ 0.00741 & COrE \\
~ non-uniform & 11.88 & 0.99 &  0.05094 $\pm$ 0.00799 & LiteBIRD \\
~ synchrotron & 65.69 & 1.00 &  0.05620 $\pm$ 0.00866 & PIXIE \\
~ spectral index & 14.96 & 1.07 &  0.05507 $\pm$ 0.00827 & EPIC-LC-TES \\
 & 15.65 & 0.98 &  0.04790 $\pm$ 0.00766 & EPIC-CS \\
 & 18.01 & 1.00 &  0.05110 $\pm$ 0.00726 & EPIC-IM-4K \\
 & 40.14 & 1.00 &  0.05075 $\pm$ 0.00745 & PRISM \\
\hline
Model 1d & 38.15 & 1.00 &  0.06756 $\pm$ 0.01027 & COrE+ Light \\
 & 42.29 & 1.01 &  0.06390 $\pm$ 0.00946 & COrE+ Extended \\
~ $r=0.05$  & 30.30 & 1.01 &  0.06074 $\pm$ 0.00920 & COrE \\
~ incorrect & 12.14 & 1.01 &  0.07988 $\pm$ 0.01027 & LiteBIRD \\
~ synchrotron & 66.47 & 1.01 &  0.07122 $\pm$ 0.01027 & PIXIE \\
~ curvature C=+0.3& 15.21 & 1.09 &  0.07769 $\pm$ 0.01029 & EPIC-LC-TES \\
 & 15.91 & 0.99 &  0.06558 $\pm$ 0.01004 & EPIC-CS \\
 & 23.43 & 1.30 &  0.06205 $\pm$ 0.00906 & EPIC-IM-4K \\
 & 44.82 & 1.12 &  0.06386 $\pm$ 0.00925 & PRISM \\
\hline
Model 1e & 38.07 & 1.00 &  0.05784 $\pm$ 0.00888 & COrE+ Light \\
 & 42.21 & 1.01 &  0.05500 $\pm$ 0.00817 & COrE+ Extended \\
~ $r=0.05$  & 30.53 & 1.02 &  0.05318 $\pm$ 0.00780 & COrE \\
~ incorrect & 12.17 & 1.01 &  0.06356 $\pm$ 0.00972 & LiteBIRD \\
~ synchrotron & 65.84 & 1.00 &  0.06119 $\pm$ 0.00939 & PIXIE \\
~ curvature C=-0.3& 15.65 & 1.12 &  0.07379 $\pm$ 0.01030 & EPIC-LC-TES \\
 & 16.04 & 1.00 &  0.05817 $\pm$ 0.00899 & EPIC-CS \\
 & 21.27 & 1.18 &  0.05267 $\pm$ 0.00750 & EPIC-IM-4K \\
 & 43.02 & 1.08 &  0.05263 $\pm$ 0.00773 & PRISM \\
\hline
\hline
Model 2c & 37.83 & 1.00 &  0.00067 $\pm$ 0.00028 & COrE+ Light \\
 & 41.68 & 0.99 &  0.00018 $\pm$ 0.00014 & COrE+ Extended \\
~ $r=0$  & 29.79 & 0.99 &  0.00021 $\pm$ 0.00016 & COrE \\
~ non-uniform & 11.88 & 0.99 &  0.00076 $\pm$ 0.00035 & LiteBIRD \\
~ synchrotron & 65.67 & 1.00 &  0.00086 $\pm$ 0.00057 & PIXIE \\
~ spectral index & 15.01 & 1.07 &  0.00059 $\pm$ 0.00038 & EPIC-LC-TES \\
 & 15.65 & 0.98 &  0.00047 $\pm$ 0.00030 & EPIC-CS \\
 & 18.01 & 1.00 &  0.00021 $\pm$ 0.00017 & EPIC-IM-4K \\
 & 40.13 & 1.00 &  0.00023 $\pm$ 0.00018 & PRISM \\
\hline
Model 2d & 38.11 & 1.00 &  0.01252 $\pm$ 0.00203 & COrE+ Light \\
 & 42.25 & 1.01 &  0.01019 $\pm$ 0.00151 & COrE+ Extended \\
~ $r=0$  & 30.27 & 1.01 &  0.01024 $\pm$ 0.00162 & COrE \\
~ incorrect & 12.19 & 1.02 &  0.02608 $\pm$ 0.00428 & LiteBIRD \\
~ synchrotron & 66.46 & 1.01 &  0.01313 $\pm$ 0.00236 & PIXIE \\
~ curvature C=+0.3& 15.24 & 1.09 &  0.01830 $\pm$ 0.00283 & EPIC-LC-TES \\
 & 15.93 & 1.00 &  0.01647 $\pm$ 0.00269 & EPIC-CS \\
 & 23.45 & 1.30 &  0.01135 $\pm$ 0.00148 & EPIC-IM-4K \\
 & 44.83 & 1.12 &  0.01211 $\pm$ 0.00156 & PRISM \\
\hline
Model 2e & 38.03 & 1.00 &  0.00499 $\pm$ 0.00107 & COrE+ Light \\
 & 42.15 & 1.00 &  0.00338 $\pm$ 0.00062 & COrE+ Extended \\
~ $r=0$  & 30.48 & 1.02 &  0.00495 $\pm$ 0.00131 & COrE \\
~ incorrect & 12.21 & 1.02 &  0.01149 $\pm$ 0.00260 & LiteBIRD \\
~ synchrotron & 65.82 & 1.00 &  0.00538 $\pm$ 0.00136 & PIXIE \\
~ curvature C=-0.3& 15.72 & 1.12 &  0.01829 $\pm$ 0.00258 & EPIC-LC-TES \\
 & 16.03 & 1.00 &  0.01080 $\pm$ 0.00224 & EPIC-CS \\
 & 21.22 & 1.18 &  0.00322 $\pm$ 0.00042 & EPIC-IM-4K \\
 & 42.98 & 1.07 &  0.00324 $\pm$ 0.00093 & PRISM \\
\hline
\end{tabular}
\caption{Maximum likelihood estimate of the tensor-to-scalar ratio $r$ (fourth column) for different experiments for sky simulations with baseline $r=0.05$ (resp. $r=0$) and $\tau = 0.087$. Model 1c: neglected variability of synchrotron spectral index, $r=0.05$. Model 1d: neglected positive synchrotron curvature $C=+0.3$, $r=0.05$. Model 1e: neglected negative synchrotron curvature $C=-0.3$, $r=0.05$. Model 2c: neglected variability of synchrotron spectral index, $r=0$ (no CMB B-modes).   Model 2d: neglected positive synchrotron curvature $C=+0.3$, $r=0$ (no CMB B-modes). Model 2e: neglected negative synchrotron curvature $C=-0.3$, $r=0$ (no CMB B-modes). In addition, mean $\chi^2$ values (second column) and normalized $\chi^2$ values (third column) from Bayesian parametric fitting for the separation of the component maps. $N_{ch}$ denotes the number of channels for each experiment (twice the number of frequencies because of $Q$ and $U$ polarization maps).}
\label{tab:results_sync}
\end{table*}

\setlength{\tabcolsep}{2mm}

Figure~\ref{Fig:masks} shows the effect of masking for the COrE experiment when incorrectly modelling the thermal dust polarization. The bias in the recovered cosmological signal $r$ is reduced when using a mask that is designed from thermal dust intensity thresholds, such as the {\it Planck}-based mask. Masking the Galactic plane $10$\,\% more conservatively leads to a significant change in the bias of the recovered tensor-to-scalar ratio. Again, a check of instability of the mean value of the tensor-to-scalar distribution when changing the Galactic mask used in the component separation process provides a robust test of spurious foreground B-mode detection and incorrect foreground modelling. 

This highlights a major advantage of the parametric pixel-fitting method employed here: in practice, the $\chi^2$ map can be used to update the model for pixels that do not provide a good fit to the data, or, alternatively, the mask can be increased to ignore those pixels thereby improving the overall result \citep{Eriksen2008}.

%%%%%%%%%%%%%%%%%%%%%%%%%%%%%%%%%%%%%%%%%%%%%%%%%%%%%%
\begin{figure*}
  \begin{center}
    \includegraphics[width=2\columnwidth]{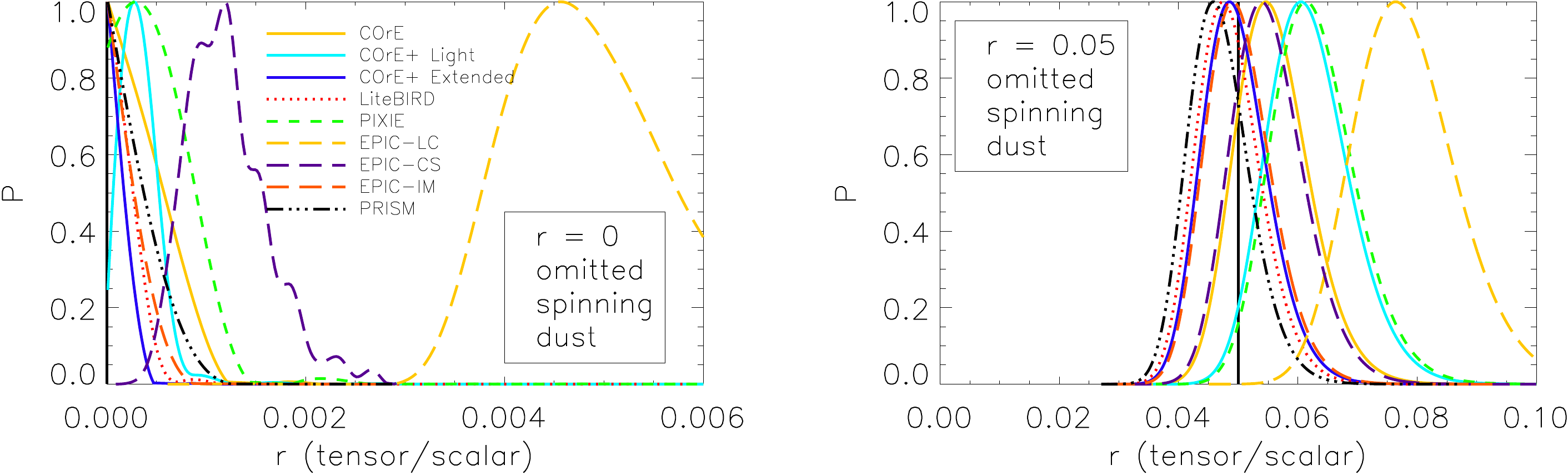}~
  \end{center}
\caption{Recovered posterior distribution $P(r)$ of the tensor-to-scalar ratio and impact of neglecting the spinning dust polarization.
The theoretical input tensor-to-scalar value (vertical solid black line) is $r=0$ in the \emph{left-hand panel} (Model 2f) and $r=0.05$ in the \emph{right-hand panel} (Model 1f). Recovered tensor-to-scalar distributions: COrE (solid yellow), COrE+ Light (solid light-blue), COrE+ Extended (solid blue), LiteBIRD (dotted red), PIXIE (dashed green), EPIC-LC-TES (long-dashed yellow), EPIC-CS (long-dashed purple), EPIC-IM-4K (long-dashed orange), PRISM (dash three-dot black).}
\label{Fig:summary_r3}
\end{figure*}
%%%%%%%%%%%%%%%%%%%%%%%%%%%%%%%%%%%%%%%%%%%%%%%%%%%%%%

\subsubsection{Incorrect spectral modelling of synchrotron}
\label{subsubsec:sync}

While polarization of the Galactic synchrotron emission is expected to scale with frequency approximately according to a power-law spectral distribution, $\nu^{\beta_s}$, with $\beta_s \approx -3$ \citep{Planck2015_XXV}, the exact value of the synchrotron spectral index $\beta_s(p,\nu)$ is certainly non-uniform over the sky \citep[e.g., ][]{Platania1998,Miville-Deschenes2008,Irfan2015} and may also vary with frequency (curvature) \citep{Kogut2007,deOliveira-Costa2008,Kogut2012}. These uncertainties on the spectral properties of the synchrotron may result in an incorrect fitting to the data at low frequency, therefore an incorrect foreground subtraction, and subsequently a biased estimation of the tensor-to-scalar ratio.

In the top panels of Fig.~\ref{Fig:summary_r2} (Models 1c and 2c), we consider a configuration where the sky data include synchrotron polarization with a non-uniform spectral index over the sky \mbox{\citep{Miville-Deschenes2008}} but we parametrize the synchrotron in the fitting model by a uniform $\beta_s = -3\pm 0.1$ spectral index (Gaussian prior). The complete set of frequency channels is used in the parametric fitting for all CMB experiments. Making the wrong assumption of a uniformly distributed synchrotron spectral index for $r=0.05$ (top right-hand panel of Fig.~\ref{Fig:summary_r2}) has a negligible impact on the recovered tensor-to-scalar ratio for all CMB satellites, the corresponding estimates of $r$ listed in the first block of Table~\ref{tab:results_sync} show consistent values with the baseline estimates of Model 1a (Table~\ref{tab:results_dust}) where the foreground model was matching the data. Therefore, at the level of sensitivity of the CMB satellites considered in this work, the non-uniform distribution over the sky of the synchrotron spectral index  \citep{Miville-Deschenes2008} is successfully fitted by the Gaussian distribution $\beta_s = -3\pm 0.1$ on $f_{\rm sky}=73\%$ of the sky (Fig.~\ref{Fig:mask_maps}).

When $r=0$ (top left-hand panel of Fig.~\ref{Fig:summary_r2}), COrE+ Light and LiteBIRD experiments recover a non-zero value of the tensor-to-scalar ratio. However, the mean $\chi^2$ of the parametric fitting (Model 2c in Table~\ref{tab:results_sync}) for these experiments show consistent values with the baseline Model 2a (correct foreground modelling), therefore indicating a good fit to the data in the frequency range considered. In this case, the bias on the tensor-to-scalar ratio is rather attributed to the lack of low-frequency channels for COrE+ Light and LiteBIRD, for which $\nu \geq 60$\,GHz. The lack of low-frequency observations at $\nu < 60$\,GHz for CMB satellites prevents them to fit for a distribution of the synchrotron spectral index that is more complex than a simple constant spectral index. As shown in top left-hand panel of Fig.~\ref{Fig:summary_r2} and the fourth block of Table~\ref{tab:results_sync}, because of the absence of low-frequency channels, LiteBIRD and COrE+ Light missions may fail in recovering the correct tensor-to-scalar ratio if the spectral distribution of the polarized synchrotron is non-trivial at low frequency. Sensitive low frequency surveys (e.g. C-BASS at 5\,GHz; \citealt{Irfan2015}) and also the lowest frequency channels of WMAP/{\it Planck} may also help. For example, for the case of LiteBIRD, including a C-BASS data point with 0.1\,mK\,beam$^{-1}$ noise  reduces the final CMB uncertainty by $\sim 50\,\%$; of course the exact value is very sensitive on the foreground model adopted.

In the middle panels (resp. bottom panels) of Fig.~\ref{Fig:summary_r2}, we show the impact on the estimated tensor-to-scalar ratio of neglecting in the foreground model a $C=+0.3$ (resp. $C=-0.3$) curvature of the synchrotron spectral index.\footnote{We adopt the convention that positive curvature corresponds to a flattening of the spectral distribution of the synchrotron emission, as given in Equation~\ref{eq:curvature}.} Neglecting a positive synchrotron curvature significantly biases the recovered tensor-to-scalar ratio by $1\sigma$ at least for all CMB experiments (middle right-hand panel of Fig.~\ref{Fig:summary_r2}). Note that the additional bias found for LiteBIRD compared to others is not due to incorrect fitting but again to the lack of low-frequency channels where synchrotron modelling is critical. Comparing the middle and bottom right-hand panels of Fig.~\ref{Fig:summary_r2} shows that neglecting a positive curvature of the synchrotron spectral index has more impact on the tensor-to-scalar estimation that omitting a negative synchrotron curvature. This can be interpreted by the fact that flattening the synchrotron distribution at increasing frequency makes the synchrotron less ``orthogonal'' to the CMB spectral distribution and makes the separation of CMB and synchrotron more challenging.

\subsubsection{Neglecting spinning dust polarization}
\label{subsubsec:spin}

\setlength{\tabcolsep}{2mm}

\begin{table*}
\centering
\begin{tabular}{l|l|l|l|l}
\hline
Model ID & Mean $\chi^2$ & Mean $\chi^2 / N_{ch}$ & Recovered $r$ & Experiment\\
\hline \hline
Model 1f & 37.89 & 1.00 &  0.05307 $\pm$ 0.00824 & COrE+ Light \\
 & 41.77 & 0.99 &  0.05159 $\pm$ 0.00767 & COrE+ Extended \\
~ $r=0.05$  & 30.63 & 1.02 &  0.05158 $\pm$ 0.00767 & COrE \\
~ omitted & 11.93 & 0.99 &  0.05178 $\pm$ 0.00791 & LiteBIRD \\
~ spinning dust &65.72 & 1.00 &  0.05606 $\pm$ 0.00861 & PIXIE \\
 & 16.81 & 1.20 &  0.06956 $\pm$ 0.01015 & EPIC-LC-TES \\
 & 16.05 & 1.00 &  0.05208 $\pm$ 0.00826 & EPIC-CS \\
 & 24.76 & 1.38 &  0.05165 $\pm$ 0.00737 & EPIC-IM-4K \\
 & 51.18 & 1.28 &  0.05191 $\pm$ 0.00762 & PRISM \\
\hline
Model 2f & 37.85 & 1.00 &  0.00044 $\pm$ 0.00025 & COrE+ Light \\
 & 41.73 & 0.99 &  0.00017 $\pm$ 0.00013 & COrE+ Extended \\
~ $r=0$  & 30.67 & 1.02 &  0.00051 $\pm$ 0.00036 & COrE \\
~ omitted & 11.93 & 0.99 &  0.00026 $\pm$ 0.00022 & LiteBIRD \\
~ spinning dust &65.70 & 1.00 &  0.00068 $\pm$ 0.00045 & PIXIE \\
 & 16.95 & 1.21 &  0.00573 $\pm$ 0.00118 & EPIC-LC-TES \\
 & 16.08 & 1.01 &  0.00155 $\pm$ 0.00051 & EPIC-CS \\
 & 24.88 & 1.38 &  0.00026 $\pm$ 0.00021 & EPIC-IM-4K \\
 & 51.29 & 1.28 &  0.00032 $\pm$ 0.00024 & PRISM \\
\hline
\end{tabular}
\caption{Maximum likelihood estimate of the tensor-to-scalar ratio $r$ (fourth column) for different experiments for sky simulations with baseline $r=0.05$ (resp. $r=0$) and $\tau = 0.087$. Model 1f: omitted spinning dust, $r=0.05$. Model 2f: omitted spinning dust, $r=0$ (no CMB B-modes). In addition, mean $\chi^2$ values (second column) and normalized $\chi^2$ values (third column) from Bayesian parametric fitting for the separation of the component maps. $N_{ch}$ denotes the number of channels for each experiment (twice the number of frequencies because of $Q$ and $U$ polarization maps). }
\label{tab:results_spin}
\end{table*}

\setlength{\tabcolsep}{2mm}

The number of polarized foregrounds in the sky is not perfectly known. Apart from thermal dust and synchrotron emissions, extra polarized components such as free-free, magnetic dust (MD), and spinning dust emissions should not be omitted. Free-free emission is probably not a major component since it is intrinsically unpolarized \citep{Rybicki_book,Keating1998,Macellari2011} while magneto-dipole radiation from magnetized dust grains could be significantly polarized, up to $\approx 35$\,\% according to some theoretical models \citep{Draine1999,Draine2013}. The latter could potentially be a major CMB polarization foreground at $\approx 70$--150\,GHz where the highest sensitivity channels are located (see Table~\ref{tab:experiments}). There has yet to be a definitive detection of MD although there are several indications that it may be contributing significantly to the total emission at frequencies $\approx 50$--$300$\,GHz \citep{Draine2012,Planck2015_Int_XXII} and thus it will need to be considered carefully when applying component separation.

The Galactic emission from spinning dust grains on the other hand is known to be significant with several clear detections at least along some sight-lines \citep{Planck2011_XX,Planck2015_X} and theoretical models suggest it should be slightly polarized at a fraction of $1$\,\% or less (see Sect.~\ref{sec:models}). Given the expected sensitivity of future CMB satellite missions, the small B-mode polarization of the spinning dust may play a non-negligible role when estimating the tensor-to-scalar ratio.

In Fig.~\ref{Fig:summary_r3} we show the impact on the tensor-to-scalar ratio of neglecting in the foreground model the $1$\% polarization of spinning dust. The sky simulations Model 1f and Model 2f  differ from the baseline simulations Model 1a and Model 2a just by the insertion of spinning dust polarization in the data. However, when omitting the $1$\,\% polarization of the spinning dust in the foreground model, the recovered tensor-to-scalar ratio is clearly impacted as it is shown by comparing the panels of Fig.~\ref{Fig:summary_r3} (Models 1f and 2f) with the second row of panels in Fig.~\ref{Fig:summary_r} (Models 1a and 2a). In particular, neglecting spinning dust polarization results in a non-negligible bias on the recovered tensor-to-scalar ratio for the EPIC experiment (also see Table~\ref{tab:results_spin}), with a large $\chi^2$ (Table~\ref{tab:results_spin}) compared to the baseline value, thus indicating imperfect component separation. Typically, neglecting the spinning dust has more impact on the CMB satellites having larger sensitivity at low frequency ($\approx 30$-$45$\,GHz), where the spinning dust polarization intensity is more significant.

%%%%%%%%%%%%%%%%%%%%%%%%%%%%%%%%%%%%%%%%%%%%%%%%%%%%
%%%
%%% Conclusions
%%%
\section{Conclusions}
\label{sec:conclusions}

The Bayesian framework for B-mode component separation allows the propagation of errors in foreground modelling towards the estimation of the tensor-to-scalar ratio. It is also useful to re-adjust, a posteriori, the model of Galactic foreground polarization being fitted by using feedback from the map of the $\chi^2$ goodness-of-fit statistics.

We have compared the overall sensitivity to B-modes of different CMB satellite concepts both in the absence and in the presence of foreground contamination. This is the first time that these proposed experiments have been compared on an equal basis.

Due to unprecedented sensitivity of next-generation CMB satellites, any incorrect spectral assumptions on the Galactic foregrounds may result in a significantly biased estimation of the tensor-to-scalar ratio after foreground cleaning. Modelling two MBB thermal dust components as a single MBB thermal dust component strongly biases the estimate of the tensor-to-scalar ratio by more than $5\sigma$ for the most sensitive experiments like PRISM, EPIC-IM, and COrE+ extended. Neglecting the synchrotron curvature, $C=0.3$, biases all CMB satellites by $1\sigma$ to $2\sigma$. Omitting the $1$\,\% polarization of spinning dust in the parametrization of Galactic foregrounds makes a non-negligible bias on the recovered tensor-to-scalar ratio. 

In this work we have used two distinct criteria to evaluate the accuracy of the CMB B-mode estimation: the $\chi^2$ statistics of the parametric fitting and the bias on the recovered tensor-to-scalar ratio. Taken together these criteria indicate whether the incorrect estimation of the tensor-to-scalar ratio results from incorrect fitting of the Galactic foregrounds or from a lack of frequency channel observations. In particular, the lack of high/low-frequency channels (e.g. \mbox{LiteBIRD}) may prevent from fitting for non-trivial foregrounds. Most important, we have shown how the lack of high/low-frequency channels and the low sensitivity in individual channels for can lead to a low $\chi^2$ value from parametric fitting with still a biased estimate of $r$, therefore preventing such setups from detecting the incorrect foreground modelling and leading to false detection of $r$. A positive curvature of the synchrotron spectral index flattens the spectrum of the synchrotron, which over a limited frequency range makes it less distinct from the flat CMB spectrum. While the fit of the global sky emission can be accurate over the restricted frequency range resulting in a low $\chi^2$, yet the CMB and the curved synchrotron are not successfully splitted, therefore leading to an extra bias on $r$. Note that an extended setup of the LiteBIRD mission including extra frequency channels and broader frequency coverage will be proposed to JAXA (Masashi Hazumi and Tomotake Matsumura, priv. comm.). Such an extension of the original LiteBIRD concept goes in the right direction of the conclusions of the paper.

We have limited the analysis to low multipoles ($\ell < 12$) where lensing B-modes are not expected to dominate the uncertainty on the tensor-to-scalar ratio. However, if $r$ is low and $\tau$ is low (in particular, latest results from \citet{2015arXiv150201589P} indicate a lower value $\tau=0.066$) then the error due to lensing B-modes, more than cosmic variance, should be considered in addition to foregrounds for a complete comparison of the satellite concepts \citep{Smith2012}. While a large panel of frequency channels is helpful for CMB satellites to clean non-trivial foreground B-modes, high-resolution channels should also be useful to improve the ability of the experiment to correct for lensing B-modes.

Given the expected sensitivity of next-generation CMB satellite missions, it appears to be fundamental to accurately characterize the spectral properties of the Galactic foregrounds in order to detect primordial CMB B-modes and to correctly measure the tensor-to-scalar ratio. In this respect, CMB satellite missions with both low-frequency (${< 60}$\,GHz) and high-frequency (${>300}$\,GHz) channels will have enough degrees of freedom to characterize the foregrounds. Sensitive foreground-dedicated experiments at low ($\lesssim 15$\,GHz) frequencies such as S-PASS \citep{Carretti2011}, C-BASS \citep{Irfan2015} and QUIJOTE \citep{Rubino-Martin2012b}, and balloon experiments at high ($\gtrsim 300$\,GHz) frequencies such as PILOT \citep{Misawa2014} will also be very helpful.

%%%%%%%%%%%%%%%%%%%%%%%%%%%%%%%%%%%%%%%%%%%%%%%%%%%%
%%%
%%% Acknowledgements
%%%
\section*{Acknowledgements}

%MR and CD acknowledge support from an ERC Starting Grant (no.~307209). 

The research leading to these results has received funding from the European Research Council under the European Union's Seventh Framework Programme (FP7/2007--2013) / ERC grant agreement no.~307209 (CD/MR). CD also acknowledges support from an STFC Consolidated Grant (no.~ST/L000768/1). HKE acknowlegdes support from the ERC Starting Grant StG2010-257080. Part of the research was carried out at the Jet Propulsion Laboratory, California Institute of Technology, under a contract with NASA. We have extensively used the {\sc HEALPix} package \citep{Gorski2005}. We acknowledge Dale Fixsen and Al Kogut for providing to us the instrumental characteristics of the PIXIE space mission. We would like to thank Jacques Delabrouille for useful comments on the draft.

\bibliographystyle{mn2e}
\bibliography{refs}

%%%%%%%%%%%%%%%%%%%%%%%%%%%%%%%%%%%%%%%%%%%%%%%%%%%%
%%%%%%%%%%%%%%%%%%%%%%%%%%%%%%%%%%%%%%%%%%%%%%%%%%%%
%%%%%%%%%%%%%%%%%%%%%%%%%%%%%%%%%%%%%%%%%%%%%%%%%%%%
%%%
%%% Appendix
%%%
%%%\newpage

\end{document}